\documentclass[a4paper,11pt]{article}
\pdfoutput=1
\usepackage[utf8x]{inputenc}
\usepackage{jheppub} 

\usepackage{amsmath,amssymb}
\usepackage{subfig}
\usepackage{slashed}
\newcommand{\Tr}[1]{\text{Tr}\big[#1\big]}

\title{Multi-component Fermionic Dark Matter and IceCube PeV scale Neutrinos in Left-Right Model with Gauge Unification}

\author[a]{Debasish Borah,}
\author[b]{Arnab Dasgupta,}
\author[c]{Ujjal Kumar Dey,}
\author[d]{Sudhanwa Patra,}
\author[e]{Gaurav Tomar}

\affiliation[a]{Department of Physics, Indian Institute of Technology Guwahati, Assam 781039, India}
\affiliation[b]{Institute of Physics, HBNI, Sachivalaya Marg, Bhubaneshwar 751005, India}
\affiliation[c]{Centre for Theoretical Studies, Indian Institute of Technology Kharagpur, Kharagpur 721302, India}
\affiliation[d]{Center of Excellence in Theoretical and Mathematical Sciences, Siksha `O' Anusandhan University, Bhubaneswar 751030, India}
\affiliation[e]{Physical Research Laboratory, Ahmedabad 380009, India}

\emailAdd{dborah@iitg.ernet.in}
\emailAdd{arnab.d@iopb.res.in}
\emailAdd{ujjal@cts.iitkgp.ernet.in}
\emailAdd{sudha.astro@gmail.com}
\emailAdd{tomar@prl.res.in}

\abstract{We consider a simple extension of the minimal left-right symmetric model (LRSM) in order to explain the PeV neutrino events seen at the IceCube experiment from a heavy decaying dark matter. The dark matter sector is composed of two fermions: one at PeV scale and the other at TeV scale such that the heavier one can decay into the lighter one and two neutrinos. The gauge annihilation cross sections of PeV dark matter are not large enough to generate its relic abundance within the observed limit. We include a pair of real scalar triplets $\Omega_{L,R}$ which can bring the thermally overproduced PeV dark matter abundance into the observed range through late time decay and consequent entropy release thereby providing a consistent way to obtain the correct relic abundance without violating the unitarity bound on dark matter mass. Another scalar field, a bitriplet under left-right gauge group is added to assist the heavier dark matter decay. The presence of an approximate global $U(1)_X$ symmetry can naturally explain the origin of tiny couplings required for long-lived nature of these decaying particles. We also show, how such an extended LRSM can be incorporated within a non-supersymmetric $SO(10)$ model where the gauge coupling unification at a very high scale naturally accommodate a PeV scale intermediate symmetry, required to explain the PeV events at IceCube.}

\begin{document}
\maketitle
\flushbottom

\section{Introduction}
\label{sec:intro}
There have been several irrefutable experimental observations suggesting the presence of dark matter (DM) in the Universe, starting from the galaxy cluster observations by Fritz Zwicky \cite{Zwicky:1933gu} back in 1933, observations of galaxy rotation curves in 1970's \cite{Rubin:1970zza}, the more recent observation of the bullet cluster \cite{Clowe:2006eq} to the latest cosmology data provided by the Planck satellite \cite{Ade:2015xua}. The latest Planck data suggest that around $26\%$ of the present Universe's energy density consist of non-baryonic or dark matter. In terms of density parameter and $h = \text{(Hubble Parameter)}/100$, the present dark matter abundance is conventionally reported as \cite{Ade:2015xua}
\begin{equation}
\Omega_{\text{DM}} h^2 = 0.1187 \pm 0.0017.
\label{dm_relic}
\end{equation}
Even though there are enough evidences from astrophysics and cosmology confirming the presence of dark matter in the Universe, no experiment has so far been able to probe it directly. The most recent dark matter direct detection experiments like LUX, PandaX-II have also reported their null results \cite{manalaysay2016, Akerib:2016vxi, Tan:2016zwf}. There have been many efforts to look for indirect dark matter signatures at different experiments with the hope that even though dark matter may not scatter off nuclei significantly in order to be consistent with null results at direct detection experiments, but they may decay or annihilate into the standard model (SM) particles on cosmological scales and leave some indirect signatures. The IceCube experiment at the south pole is one such place where such indirect dark matter signatures can be looked for. 
The IceCube collaboration has in fact reported 54 ultra-high energy (UHE) neutrino events corresponding to deposited energy in the range 20 TeV to 2 PeV in their 4 year dataset \cite{Aartsen:2013jdh, Aartsen:2014gkd, Aartsen:2015zva}, with a $6.4\sigma$ excess over the atmospheric background. Recently, IceCube has observed a track-like event which extends its deposited energy to $\sim$ 3 PeV~\cite{Aartsen2016}. With no significant evidence for astrophysical sources \cite{Adrian-Martinez:2015ver, Aartsen:2016tpb}, this observation has led to several beyond standard model (BSM) proposals as possible interpretation of these high energy neutrino events. Among them, probably the simplest BSM interpretation of these high energy events is in terms of a PeV scale long-lived dark matter candidate that can decay into neutrinos \cite{Feldstein:2013kka, Esmaili:2013gha, Ema:2013nda, Bhattacharya:2014vwa, Rott:2014kfa, Esmaili:2014rma, Bhattacharya:2014yha, Cherry:2014xra, Fong:2014bsa, Murase:2015gea, Boucenna:2015tra, Ko:2015nma, ElAisati2015, Dev:2016qbd, Bhattacharya:2016tma, Chianese:2016opp, DiBari:2016guw, Chianese:2016smc, Bai:2013nga, Higaki:2014dwa,  Ema:2014ufa, Dudas:2014bca, Kopp:2015bfa, Fiorentin:2016avj, Bhattacharya:2017jaw}. Among these, we are particularly interested in the framework adopted by \cite{Dev:2016qbd, Fiorentin:2016avj}. The authors in both of these works considered a type of Left-Right Symmetric Model with relevant field content to generate the high energy IceCube events from the decay of a PeV dark matter candidate. The LRSM is one of most highly motivated BSM frameworks which in its generic form \cite{Pati:1974yy, Mohapatra:1974gc, Senjanovic:1975rk, Mohapatra:1980qe, Deshpande:1990ip}, not only explains the origin of parity violation in weak interactions but also explains the origin of tiny neutrino masses naturally. The gauge symmetry group and the field content of the generic LRSM can also be embedded within grand unified theory (GUT) symmetry groups like $SO(10)$ providing a non-supersymmetric route to gauge coupling unification. The right handed fermions of the SM forms doublet under a new $SU(2)_R$ group in LRSM such that the theory remains parity symmetric at high energy. This necessitates the inclusion of the right handed neutrino as a part of the right handed lepton doublet. Both the works \cite{Dev:2016qbd, Fiorentin:2016avj} considered the lightest right handed neutrino as the PeV dark matter candidate. However, in generic LRSM, such a PeV right handed neutrino will be extremely short lived compared to the age of the Universe, due to its decay into the SM fermions mediated by the $SU(2)_R$ vector bosons. Therefore the authors of \cite{Dev:2016qbd, Fiorentin:2016avj} considered two different types of LRSM which can explain the IceCube result well but not as motivating as the generic LRSM from grand unification point of view. For example, the work \cite{Dev:2016qbd} considered the charged isospin partner of the right handed neutrino to be a super-heavy fermion instead of the usual right handed charged lepton, in order to prevent the fast decay of dark matter into the SM particles. On the other hand, the authors of \cite{Fiorentin:2016avj} considered a hadrophobic $SU(2)_R$ to prevent the $W_R$ mediated fast decay of the dark matter candidate. 
Instead of pursuing this non-conventional LRSM route taken in \cite{Dev:2016qbd, Fiorentin:2016avj}, here we consider a simple extension of the minimal LRSM to incorporate the IceCube observation while retaining other generic features of LRSM including the possibility of embedding it within $SO(10)$ GUT. The possibility of dark matter within GUT has been explored quite extensively within supersymmetric frameworks. Within non-supersymmetric $SO(10)$ GUT also, similar studies have appeared in some recent works including \cite{Mambrini:2015vna, Nagata:2015dma}. The interesting feature of these scenarios is the natural origin of the symmetry stabilising the dark matter candidate within a UV complete theory that can achieve gauge coupling unification and can also predict the proton decay lifetime. While we do not follow a general top-down approach here similar to \cite{Mambrini:2015vna, Nagata:2015dma}, we start with a left-right symmetric model having necessary particle content to produce the correct dark matter relic abundance along with the IceCube high energy neutrino events and then consider the possible embedding of these particles within $SO(10)$ multiplets. This approach is similar to the bottom-up approach followed in LRSM dark matter studies with gauge coupling unification discussed in the recent works \cite{Borah:2016ees, Bandyopadhyay:2017uwc, Arbelaez:2017ptu}. In our model, the neutral components of two fermion triplets are dark matter candidates with the heavier of them having PeV scale mass and the lighter one having mass at TeV scale. Although the abundance of TeV scale DM can be kept within the observed DM abundance in the usual thermal freeze-out mechanism, the PeV scale DM becomes over-abundant due to insufficient annihilations or the unitarity bound \cite{Griest:1989wd, Nussinov:2014qva}. We introduce two different types of scalar multiplets: a pair of triplet and one bi-triplet which serve two purposes. One of them can be long lived and decay after the freeze-out of PeV DM so that its abundance can be diluted by the release of entropy \cite{Scherrer:1984fd} while the other can assist the PeV DM decay into the lighter one and a pair of high energy neutrinos. The long-lived nature of such scalars as well as the heavier DM requires arbitrary fine-tuning of different parameters of the model. However, the presence of an approximate global $U(1)_X$ symmetry can naturally explain the origin of such tiny couplings required for long-lived nature of these decaying particles.

We show that the additional fields included to the minimal LRSM in order to accommodate a heavy long lived dark matter decaying into the high energy neutrinos observed at the IceCube experiment also assist in achieving successful gauge coupling unification at a high scale. Apart from the usual LRSM particles, the additional fields included in this work also can be accommodated within different $SO(10)$ multiplets. The gauge coupling unification at a high energy scale is consistent with an intermediate scale around PeV scale. Although this could well be a coincidence, but it could also be an indication suggesting a more fundamental origin of a dark matter candidate having mass around a few PeV. Interestingly, the intermediate Pati-Salam symmetry in generic $SO(10)$ GUT models has a lower bound from phenomenological considerations of rare meson decays, which lie near the PeV scale \cite{Valencia:1994cj}. Also, such PeV scale intermediate left-right symmetry has other cosmological motivations \cite{Yajnik:2017yhd}. Therefore, the PeV dark matter in our model is not just motivated from phenomenological considerations, but it could also have a deeper theoretical origin as this PeV intermediate scale arises naturally from the demand of successful gauge coupling unification and consistent cosmology.
This paper is organised as follows. In section \ref{sec:model}, we discuss our model in details. We discuss the gauge coupling unification along with $SO(10)$ embedding in section \ref{sec:unific}. In section \ref{sec:relic}, we discuss the details of dark matter relic abundance calculation and constraints on the parameter space from the requirement of satisfying the relic density bound.  In section \ref{sec:ic}, we discuss the way the IceCube high energy neutrino events can be explained from PeV dark matter decay in our model. We finally conclude in section \ref{sec:concl}.

\section{The Model}
\label{sec:model}
The left-right symmetric model \cite{Pati:1974yy, Mohapatra:1974gc, Senjanovic:1975rk, Mohapatra:1980qe, Deshpande:1990ip} is one of the most widely studied BSM framework that can simultaneously explain the origin of tiny neutrino masses and parity violation at weak interactions. The gauge symmetry of the standard model namely, $SU(3)_c \times SU(2)_L \times U(1)_Y$ is upgraded to $SU(3)_c \times SU(2)_L \times SU(2)_R \times U(1)_{B-L}$ such that the right handed fermions transform as doublets under $SU(2)_R$, making the theory left-right symmetric. The model also has an in-built discrete $Z_2$ symmetry or D-parity which ensures the equality of couplings in $SU(2)_{L,R}$ sectors. The effective parity violating electroweak physics at low energy arises as a result of spontaneous breaking of the $SU(2)_R \times U(1)_{B-L} \times D$ to $U(1)_Y$ of the SM. 
In the manifest left-right symmetric model $SU(3)_c \times SU(2)_L \times SU(2)_R \times U(1)_{B-L}$, the scale of $SU(2)_R$ gauge symmetry breaking and parity breaking are identical which is not necessary. The original idea of D-parity invariance in left-right symmetric models is proposed in~\cite{Chang:1983fu, Chang:1984uy, Chang:1984qr} where the discrete parity symmetry gets broken much before the $SU(2)_R$ gauge symmetry breaking. 
The key difference between Lorentz parity and D-parity is that Lorentz parity acts on the Lorentz group and interchanges left-handed fermions with the right-handed ones but the bosonic fields remain the same whereas D-parity acts on the gauge groups $SU(2)_L \times SU(2)_R$ interchanging the $SU(2)_L$ scalar fields with the $SU(2)_R$ scalar fields in addition to the interchange of fermions. The effect of the spontaneous breaking of D-parity results in an asymmetry between left and right-handed scalar fields making the coupling constants of $SU(2)_R$ and $SU(2)_L$ evolve separately under the renormalization group leading to unequal $SU(2)_L$ and $SU(2)_R$ gauge couplings. To illustrate the idea, consider the scalar sector of the left-right model with spontaneous D-parity breaking mechanism consists of a $SU(2)$ singlet scalar field $\sigma$ which is odd under discrete D-parity, two scalar triplets $\Delta_L, \Delta_R$ and a bidoublet $\Phi$.
After assigning a non-zero vacuum expectation value (vev), $\langle \sigma \rangle$ to D-parity odd scalar singlet $\sigma$, the left-right symmetry with D-parity $SU(3)_c \times SU(2)_L \times SU(2)_R \times U(1)_{B-L} \times D$ is spontaneously broken but the gauge symmetry $SU(3)_c \times SU(2)_L \times SU(2)_R \times U(1)_{B-L}$ remains unbroken resulting in 
\begin{subequations}
\begin{eqnarray}
&&M^2_{\Delta_R}=\mu^2_{\Delta} - \lambda_{\Delta} \langle \sigma \rangle M\, ,  \\
&&M^2_{\Delta_L}=\mu^2_{\Delta} + \lambda_{\Delta} \langle \sigma \rangle M\, ,
\end{eqnarray}
\end{subequations}
where $\mu_\Delta$ is the mass term for triplets i.e, $\mu^2_{\Delta} \mbox{Tr}\left(\Delta^\dagger_L \Delta_L+ 
\Delta^\dagger_R \Delta_R\right)$ and $\lambda_\Delta$ is the trilinear coupling in the term $M \sigma \mbox{Tr}\left(\Delta^\dagger_L \Delta_L - \Delta^\dagger_R \Delta_R\right)$.
With a large parity breaking vev to the D-parity odd scalar singlet; $\langle \sigma \rangle \sim M_{\text{Pl}}$, the model yields all the left-handed (LH) scalars to have heavy masses i.e., $\mathcal{O}(M_{\text{Pl}})$ while those of the right-handed (RH) scalars can have much lighter masses near the TeV scale with $M^2_{\Delta_R} \simeq \left(\mu^2_{\Delta} -\lambda_{\Delta} \langle \sigma\rangle M \right)$ where $\mu_{\Delta} \sim M \sim \mathcal{O} (M_{\text{Pl}})$. In fact $M_{\Delta_R}$ can have any value below $M_{\text{Pl}}$ depending on the degree of fine-tuning in the scalar coupling $\lambda_{\Delta}$. Thus, the spontaneous D-parity breaking is not only gives mass splitting between left and right handed scalar fields, but the asymmetry in the scalar sector at the energy scales below D-parity breaking scale $M_{\text{Pl}}$ creates an asymmetry in the gauge couplings, $g_{L} \neq g_{R}$ for the surviving left-right gauge group. We implement this nice idea in the present model to provide required mass splitting between Dark matter fermions as well as scalars.
The usual fermion content of the LRSM along with the dark matter candidates $\Sigma_{L,R}$ added in the present model are shown in table \ref{tab1}. The scalar content of the minimal LRSM consist of a bidoublet $\Phi$ responsible for generating Dirac mass terms of all fermions and also for breaking the electroweak gauge symmetry spontaneously. The minimal model also has a pair of complex triplet scalars $\Delta_{L,R}$ in order to break the LRSM gauge symmetry spontaneously to that of the SM and also to generate Majorana mass terms of light and heavy neutrinos. The scalar content of the present model with the addition of a pair of real triplets, one bitriplet and one D-parity odd singlet, is shown in table \ref{tab2}. The parity odd scalar not only helps in gauge coupling unification as we will discuss later, but also splits the masses of left and right handed dark matter candidates.
\begin{table}
\begin{center}
\begin{tabular}{|c|c|}
\hline
Particles & $SU(3)_c \times SU(2)_L \times SU(2)_R \times U(1)_{B-L}$   \\
\hline
$Q_L$ & $(3,2,1,\frac{1}{3})$ \\
$Q_R$ & $(3,1,2,\frac{1}{3})$  \\
$\ell_L$ & $(1,2,1,-1)$  \\
$\ell_R$ & $(1,1,2,-1)$ \\
$\Sigma_L$ & $(1,3,1,0)$ \\
$\Sigma_R$ & $(1,1,3,0)$ \\
\hline
\end{tabular}
\end{center}
\caption{Fermion content of the model}
\label{tab1}
\end{table}
\begin{table}
\begin{center}
\begin{tabular}{|c|c|}
\hline
Particles & $SU(3)_c \times SU(2)_L \times SU(2)_R \times U(1)_{B-L}$   \\
\hline
$\Phi$ & $(1,2,2,0)$ \\
$\Delta_L$ & $(1,3,1,2)$ \\
$\Delta_R$ & $(1,1,3,2)$ \\
$\Omega_L$ & $(1,3,1,0)$ \\
$\Omega_R$ & $(1,1,3,0)$ \\
$\psi$ & $(1,3,3,0)$\\
$ \sigma  $ & $(1,1,1,0)$\\
\hline
\end{tabular}
\end{center}
\caption{Scalar content of the model}
\label{tab2}
\end{table}

At a very high energy scale, the parity odd singlet $\sigma$ can acquire a vev to break D-parity spontaneously while the neutral component of $\Delta_R$ acquires a non-zero vev at a later stage to break the gauge symmetry of the LRSM into that of the SM which then finally gets broken down to the $U(1)_{\rm em}$ of electromagnetism by the vev of the neutral component of Higgs bidoublet $\Phi$. Thus, the symmetry breaking chain is 
$$ SU(2)_L \times SU(2)_R \times U(1)_{B-L} \times D \quad \underrightarrow{\langle \sigma \rangle} \quad SU(2)_L \times SU(2)_R \times U(1)_{B-L} \quad \underrightarrow{\langle
\Delta_R \rangle} \quad SU(2)_L\times U(1)_Y$$
$$ SU(2)_L\times U(1)_Y \quad \underrightarrow{\langle \Phi \rangle} \quad U(1)_{\rm em}$$
Denoting the vev of the neutral components of the bidoublet as $k_{1,2}$ and that of triplet $\Delta_R$ as $v_R$, the gauge boson masses after spontaneous symmetry breaking can be written as
$$ M^2_{W_L} = \frac{g^2}{4} k^2_1, \;\;\; M^2_{W_R} = \frac{g^2}{2}v^2_R ~, $$
$$ M^2_{Z_L} =  \frac{g^2 k^2_1}{4\cos^2{\theta_w}} \left ( 1-\frac{\cos^2{2\theta_w}}{2\cos^4{\theta_w}}\frac{k^2_1}{v^2_R} \right), \;\;\; M^2_{Z_R} = \frac{g^2 v^2_R \cos^2{\theta_w}}{\cos{2\theta_w}}~, $$
where $\theta_w$ is the Weinberg angle. The neutral components of the other scalar fields $\psi, \Omega_{L,R}$ do not acquire any vev. However, the neutral component of the scalar triplet $\Delta_L$ can acquire a tiny but non-zero induced vev after the electroweak symmetry breaking as
\begin{equation}
v_{L}=\gamma \frac{M^{2}_{W_L}}{v_{R}}~,
\label{deltaLvev}
\end{equation}
with $M_{W_L}\sim 80.4$ GeV being the weak boson mass and $\gamma$ is a function of various couplings in the scalar potential. The bidoublet also gives rise to non-zero $W_L-W_R$ mixing parameterised by $\xi$ as
\begin{equation}
\tan{2\xi} = \frac{2k_1 k_2}{v^2_R-v^2_L}\;,
\end{equation}
which is constrained to be $\xi \leq 7.7 \times 10^{-4}$ \cite{ATLAS:2012ak, CMS:2012zv}. Such tiny left-right mixing simplifies the calculation of dark matter relic abundance as we discuss in detail later.

The relevant Yukawa couplings for the standard model fermion masses can be written as 
\begin{align}
{\cal L}^{SM}_{Y} &= y_{ij} \bar{\ell}_{iL} \Phi \ell_{jR}+ y^\prime_{ij} \bar{\ell}_{iL}
\tilde{\Phi} \ell_{jR} +Y_{ij} \bar{Q}_{iL} \Phi Q_{jR}+ Y^\prime_{ij} \bar{Q}_{iL}
\tilde{\Phi} Q_{jR}
\nonumber \\
& \qquad + \frac{1}{2}(f_L)_{ij} \ell_{iL}^T \ C \ i \sigma_2 \Delta_L \ell_{jL}+\frac{1}{2}(f_R)_{ij} \ell_{iR}^T \ C \ i \sigma_2 \Delta_R \ell_{jR}+\text{H.c.}
\label{treeY}
\end{align}
where $\tilde{\Phi} = \tau_2 \Phi^* \tau_2$, $C$ is the charge conjugation operator and the indices $i, j = 1, 2, 3$ correspond to the three generations of fermions. The light neutrino mass after the spontaneous symmetry breaking can be written as 
\begin{equation}
(M_{\nu})_{ij} = (f_L)_{ij} v_L - (M_D)_{ik} (M_R)^{-1}_{kl} (M^T_D)_{lj}
\end{equation}
where $(M_D)_{ij} = y_{ij}k_1 + y^\prime_{ij} k_2$ are the elements of Dirac neutrino mass matrix and $(M_R)_{ij} = (f_R)_{ij} v_R$ are the right handed neutrino mass matrix elements.

The fermion triplets $\Sigma_{L,R}$ shown in table \ref{tab1} can be written in the component form as the following matrix representation 
\begin{eqnarray}
&&\Sigma_{L}=\begin{pmatrix}
  \Sigma^0_{L}  & \sqrt{2} \Sigma^+_{L}  \\
  \sqrt{2} \Sigma^-_{L} & -\Sigma^0_{L}
 \end{pmatrix}, \nonumber \\
&&\Sigma_{R}=\begin{pmatrix}
  \Sigma^0_{R}  & \sqrt{2} \Sigma^+_{R}  \\
  \sqrt{2} \Sigma^-_{R} & -\Sigma^0_{R}
 \end{pmatrix}, 
\end{eqnarray}
where the neutral component of the each fermion triplet can be a stable dark matter candidate whose stability is ensured by either high $SU(2)$ dimensions forbidding any decay at tree level or due to remnant discrete symmetry arising after spontaneous symmetry breaking of LRSM~\cite{Heeck:2015qra}. There exists a parity odd singlet scalar $\sigma$ that can introduce a mass splitting between $\Sigma_L$ and $\Sigma_R$. The terms involving $\Sigma_{L,R}$ and $\sigma$ can be written as
\begin{equation}
{\cal L}^{\Sigma \sigma}_{Y} = M_L \Sigma^T_L C \Sigma_L + M_R\Sigma^T_R C \Sigma_R+\lambda_{\sigma} \sigma (\Sigma_R \Sigma_R-\Sigma_L \Sigma_L).
\label{sigmasigma}
\end{equation}
A non-zero vev of the parity odd singlet scalar $\sigma$ lead to the masses of fermion triplets as
$$M_R = M_{\Sigma} + \lambda_{\sigma} \langle \sigma \rangle, \;\; M_L = M_{\Sigma} - \lambda_{\sigma} \langle \sigma \rangle~.$$
This generates a mass splitting of $2\lambda_{\sigma} \langle \sigma \rangle$ between the two dark matter particles. Since we want our heavier dark matter candidate $\Sigma_R$ to decay into the lighter one and other standard model particles, we introduce a scalar bitriplet into the model which can mediate such decays. Such a scalar bitriplet field $\psi \sim (1, \textbf{3},\textbf{3},0)$ can be written in matrix form as
\begin{equation}
\psi = \left(\begin{array}{ccc}
\zeta^{0*} & \epsilon^+ & \zeta^{++} \\
-\zeta^{+*} & \epsilon^0 & \zeta^{+} \\
\zeta^{++*} & -\epsilon^{+*} & \zeta^0 \\
\end{array}
\right).
\end{equation}
The relevant interaction Lagrangian is given by
\begin{align}
\mathcal{L}^{\Sigma \psi}_Y= \frac{f_{\psi}}{2} \overline{\Sigma_L} \psi \Sigma_R
\label{psisigma}
\end{align}
where the contribution of the parity odd singlet in the desired mass splitting $M_R \neq M_L$ is already taken into account. A pair of scalar triplets $\Omega_{L,R}$ is introduced in order to achieve the desired relic abundance of the heavier dark matter candidate as we discuss below. These scalar triplets couple to the fermion dark matter triplets as 
\begin{align}
\mathcal{L}^{\Sigma \Omega}_Y = \lambda \Sigma^T_L C \Omega_L \Sigma_L + \lambda\Sigma^T_R C \Omega_R \Sigma_R\;.
\label{sigmaomega}
\end{align}
The complete Yukawa Lagrangian of the model therefore, can be written as
\begin{equation}
\mathcal{L}_Y = \mathcal{L}^{SM}_Y + {\cal L}^{\Sigma \sigma}_{Y} + \mathcal{L}^{\Sigma \psi}_Y+\mathcal{L}^{\Sigma \Omega}_Y
\end{equation}
where $\mathcal{L}^{SM}_Y, {\cal L}^{\Sigma \sigma}_{Y}, \mathcal{L}^{\Sigma \psi}_Y, \mathcal{L}^{\Sigma \Omega}_Y$ can be read from Eqs.~\eqref{treeY}, \eqref{sigmasigma}, \eqref{psisigma} and \eqref{sigmaomega} respectively. For the sake of completeness we also write down the scalar potential of the model by dividing it into two parts: one for minimal LRSM and the other for the relevant part of the additional scalar content $(\Omega_{L,R}, \psi, \sigma)$. The scalar potential for the minimal LRSM is 
\begin{equation}
 V(\Phi,\Delta_L,\Delta_R) = V_{\mu} + V_{\Phi} + V_{\Delta} + V_{\Phi\Delta} + V_{\Phi\Delta_L\Delta_R},
\end{equation}
where the bilinear terms in Higgs fields are
\begin{eqnarray}
 V_{\mu} &=& -\mu_1^2 \Tr{\Phi^\dagger\Phi}
 - \mu_2^2\Tr{\Phi^\dagger\tilde{\Phi} + \tilde{\Phi}^\dagger\Phi}
 - \mu_3^2\Tr{\Delta_L^\dagger\Delta_L + \Delta_R^\dagger\Delta_R}.
 \end{eqnarray}
The self-interaction terms of $\Phi$ are: 
\begin{align}
 V_{\Phi} = ~
  & \lambda_1\left[\Tr{\Phi^\dagger\Phi}\right]^2 +
  \lambda_2\left[\Tr{\Phi^\dagger\tilde{\Phi}}\right]^2 + \lambda_2\left[\Tr{\tilde{\Phi}^\dagger\Phi}\right]^2 
  \nonumber \\   
  & + \lambda_3\Tr{\Phi^\dagger\tilde{\Phi}}\Tr{\tilde{\Phi}^\dagger\Phi} + 
  \lambda_4\Tr{\Phi^\dagger\Phi}\Tr{\Phi^\dagger\tilde{\Phi} + \tilde{\Phi}^\dagger\Phi}.
  \label{appeneq1}
\end{align}
and the $\Delta_{L,R}$ self- and cross-couplings are as follows:
\begin{align}
 V_{\Delta} = ~ 
 & \rho_1\left(\left[\Tr{\Delta_L^\dagger\Delta_L}\right]^2 + 
     \left[\Tr{\Delta_R^\dagger\Delta_R}\right]^2\right) +
 \rho_3\Tr{\Delta_L^\dagger\Delta_L}\Tr{\Delta_R^\dagger\Delta_R}
 \nonumber \\ 
 & + \rho_2 \left(\Tr{\Delta_L\Delta_L}\Tr{\Delta_L^\dagger\Delta_L^\dagger} + \Tr{\Delta_R\Delta_R}\Tr{\Delta_R^\dagger\Delta_R^\dagger} \right)
\nonumber  \\ 
 & + \rho_4\left(\Tr{\Delta_L\Delta_L}\Tr{\Delta_R^\dagger\Delta_R^\dagger} + \Tr{\Delta_L^\dagger\Delta_L^\dagger}\Tr{\Delta_R\Delta_R}\right).
   \label{appeneq2}
\end{align}
In addition, there are also  $\Phi-\Delta_L$ and $\Phi-\Delta_R$ interactions present in the model, 
\begin{align}
 V_{\Phi\Delta} = ~
 & \alpha_1\Tr{\Phi^\dagger\Phi}\Tr{\Delta_L^\dagger\Delta_L + \Delta_R^\dagger\Delta_R} +
 \alpha_3 \Tr{\Phi\Phi^\dagger\Delta_L\Delta_L^\dagger + \Phi^\dagger\Phi\Delta_R\Delta_R^\dagger} 
  \nonumber \\ 
 & + \left\{
 \alpha_2 e^{i\delta_2}\Tr{\Phi^\dagger\tilde{\Phi}}\Tr{\Delta_L^\dagger\Delta_L} + 
 \alpha_2 e^{i\delta_2}\Tr{\tilde{\Phi}^\dagger\Phi}\Tr{\Delta_R^\dagger\Delta_R} + \text{H.c.}\right\}
  \label{appeneq3}
\end{align}
with $\delta_2 =0$ making CP conservation explicit, and the $\Phi-\Delta_L-\Delta_R$ couplings are
\begin{align}
 V_{\Phi\Delta_L\Delta_R} = ~
 & \beta_1\Tr{\Phi^\dagger\Delta_L^\dagger\Phi\Delta_R + \Delta_R^\dagger\Phi^\dagger\Delta_L\Phi} +
 \beta_2\Tr{\Phi^\dagger\Delta_L^\dagger\tilde{\Phi}\Delta_R + \Delta_R^\dagger\tilde{\Phi}^\dagger\Delta_L\Phi}
\nonumber \\ 
& + \beta_3\Tr{\tilde{\Phi}^\dagger\Delta_L^\dagger\Phi\Delta_R + \Delta_R^\dagger\Phi^\dagger\Delta_L\tilde{\Phi}}.
\end{align}
The scalar potential involving the newly introduced scalar fields beyond the minimal LRSM is
\begin{equation}
V_{\rm new} = V_{\Omega} + V_{\psi} + V_{\sigma} + V_{\Phi\Omega}+ V_{\Delta \Omega}+V_{\psi \Omega}.
\end{equation}
The details of different terms on the right-hand side of the above equation can be written as follows,
\begin{align}
V_{\Omega} = ~ & \mu^2_{\Omega} \Tr{\Omega_L^T\Omega_L + \Omega_R^T\Omega_R} + \rho_5 (\left[\Tr{\Omega_L^T\Omega_L}\right]^2 +\left[\Tr{\Omega_R^T\Omega_R}\right]^2) \nonumber \\
& +\rho_6 \left[\Tr{\Omega_L^T\Omega_L}\right]\left[\Tr{\Omega_R^T\Omega_R}\right],
\end{align}
\begin{align}
V_{\psi} = \mu^2_{\psi} \Tr{\psi^T\psi} + \rho_7 \left[\Tr{\psi^T\psi}\right]^2,
\end{align}
\begin{align}
V_{\sigma}  = ~ & \frac{\mu^2_{\sigma}}{2} \sigma^2 + \rho_8  \sigma^4 + \mu_{\sigma \Delta} \sigma (\Tr{\Delta_R^\dagger\Delta_R - \Delta_L^\dagger\Delta_L}) +\mu_{\sigma \Omega} \sigma (\Tr{\Omega_R^T\Omega_R - \Omega_L^T\Omega_L})  \nonumber \\
& +\rho_9 \sigma^2 \Tr{\Phi^\dagger\Phi}+\rho_{10} \sigma^2 (\Tr{\Delta_R^\dagger\Delta_R + \Delta_L^\dagger\Delta_L}) + \rho_{11} \sigma^2 (\Tr{\Omega_R^T\Omega_R + \Omega_L^T\Omega_L}) \nonumber \\
&  + \rho_{12} \sigma^2 \Tr{\psi^T\psi},
\end{align}
\begin{align}
 V_{\Phi\Omega}  = ~
& \mu_{14}\Tr{\Phi^\dagger \Omega_L \Phi} + \mu_{15}\Tr{\Phi^\dagger \Omega_R \Phi} +f_{145} \Tr{\Phi^{\dagger} \Phi} \Tr{\Omega^{T}_L \Omega_R} \nonumber \\
 & + f_{14} \Tr{\Phi^{\dagger} \Phi} \Tr{\Omega^{T}_L \Omega_L}+ f_{15} \Tr{\Phi^{\dagger} \Phi} \Tr{\Omega^{T}_R \Omega_R},
 \label{phiomega}
 \end{align}
 \begin{align}
 V_{\Delta \Omega}  = ~ & f_{24} \Tr{\Delta^{\dagger}_L \Delta_L} \Tr{\Omega^{T}_L \Omega_L}+f_{25} \Tr{\Delta^{\dagger}_L \Delta_L} \Tr{\Omega^{T}_R \Omega_R}  + f_{34} \Tr{\Delta^{\dagger}_R \Delta_R} \Tr{\Omega^{T}_L \Omega_L} \nonumber \\
 & +f_{35} \Tr{\Delta^{\dagger}_R \Delta_R} \Tr{\Omega^{T}_R \Omega_R}+ f_{2345} \Tr{\Delta^{\dagger}_L \Delta_R} \Tr{\Omega^{T}_L \Omega_R},
 \label{deltaomega}
 \end{align}
\begin{align}
V_{\psi \Omega}  = ~ & \mu_{16} \Tr{\Phi^{\dagger} \psi \Phi} + \mu_{236} \Tr{\Delta^{\dagger}_L \psi \Delta_R} +\mu_{456} \Tr{\Omega^{T}_L \psi \Omega_R} + f_{16} \Tr{\Phi^{\dagger} \Phi} \Tr{\psi^{T} \psi} \nonumber \\
& + f_{26} \Tr{\Delta^{\dagger}_L \Delta_L} \Tr{\psi^{T} \psi} + f_{36}\Tr{\Delta^{\dagger}_R \Delta_R} \Tr{\psi^{T} \psi} + f_{236} \Tr{\Delta^{\dagger}_L \Delta_R} \Tr{\psi^{T} \psi} \nonumber \\
& + f_{46} \Tr{\Omega^{T}_L \Omega_L} \Tr{\psi^{T} \psi} + f_{56} \Tr{\Omega^{T}_R \Omega_R} \Tr{\psi^{T} \psi} + f_{456} \Tr{\Omega^{T}_L \Omega_R} \Tr{\psi^{T} \psi} \nonumber \\
& +\mu_{17} \Tr{\psi \psi \psi} \;.
\label{psiomega}
\end{align}
It should be noted in the scalar potential terms involving the scalar bitriplet, both the two and three dimensional representations of $SU(2)$ generators have to be used at appropriate places in order to construct an invariant term. The details of this is skipped as these terms are shown for the sake of completeness here and are not relevant for subsequent discussions. The complete scalar potential of the model is, therefore
\begin{equation}
V=  V(\Phi,\Delta_L,\Delta_R)+V_{\Omega} + V_{\psi} + V_{\sigma} + V_{\Phi\Omega}+ V_{\Delta \Omega}+V_{\psi \Omega}.
\end{equation}
On the other hand, the kinetic Lagrangian of the model can be written as
\begin{align}
\mathcal{L}_{\rm kin} = \mathcal{L}^{Q, \ell}_{\rm kin} + \mathcal{L}^{\Sigma}_{\rm kin} + \mathcal{L}^{\rm Scalar}_{\rm kin}
\end{align}
where
\begin{align}
 \mathcal{L}^{Q, \ell}_{\rm kin} = ~ & \sum_{\chi = \ell, Q} \bar{\chi}_L \gamma^{\mu} \left (i \partial_{\mu} + g_L \frac{\vec{\tau}}{2} \cdot \vec{W}_{L\mu} + g^{\prime} \frac{B-L}{2} B_{\mu} \right) \chi_L  \nonumber \\
 &+\sum_{\chi = \ell, Q} \bar{\chi}_R \gamma^{\mu} \left (i \partial_{\mu} + g_R \frac{\vec{\tau}}{2} \cdot \vec{W}_{R\mu} + g^{\prime} \frac{B-L}{2} B_{\mu} \right) \chi_R,
  \end{align}
 \begin{align}
\mathcal{L}^{\Sigma}_{\rm kin} = ~ &\Tr{ \bar{\Sigma}_L \gamma^{\mu} \left(i\partial_{\mu} \Sigma_L +g_L \big[ \frac{\vec{\tau}}{2} \cdot \vec{W}_{L \mu}, \Sigma_L \big]\right)} \nonumber \\
&+ \Tr{ \bar{\Sigma}_R \gamma^{\mu} \left(i\partial_{\mu} \Sigma_R +g_R \big[ \frac{\vec{\tau}}{2} \cdot \vec{W}_{R \mu}, \Sigma_R \big]\right)},
\end{align}
\begin{align}
\mathcal{L}^{\rm Scalar}_{\rm kin} = ~ & \Tr{(D_{\mu} \Phi)^{\dagger} (D^{\mu}\Phi)}+\Tr{(D_{\mu} \Delta_L)^{\dagger} (D^{\mu}\Delta_L)}+\Tr{(D_{\mu} \Omega_L)^{\dagger} (D^{\mu}\Omega_L)} \nonumber \\
&+\Tr{(D_{\mu} \psi)^{\dagger} (D^{\mu}\psi)} +\Tr{(D_{\mu} \Delta_R)^{\dagger} (D^{\mu}\Delta_R)}+\Tr{(D_{\mu} \Omega_R)^{\dagger} (D^{\mu}\Omega_R)} \nonumber \\
&+\frac{1}{2}(\partial_{\mu} \sigma)(\partial^{\mu} \sigma).
\end{align}
In $\mathcal{L}^{\rm Scalar}_{\rm kin}$ the covariant derivatives $D_{\mu}$ for the scalars can be written in a way similar to that of the fermions.

\section{Gauge Coupling Unification}
\label{sec:unific}
In the previous discussion, we demonstrated that PeV scale decaying dark matter has a potential to explain the recently observed IceCube data. We intend here to examine whether such a framework under consideration can be embedded in a non-SUSY $SO(10)$ GUT theory leading to successful gauge coupling unification. The symmetry breaking pattern of $SO(10)$ such that it has the Pati-Salam gauge group as its intermediate symmetry breaking step as follows
\begin{align}
	SO(10) \mathop{\longrightarrow}^{\langle \eta \rangle} G_{224D} 
               \mathop{\longrightarrow}^{\langle \sigma \rangle} G_{224} 
		   \mathop{\longrightarrow}^{\langle \Delta_R    \rangle} G_\text{SM} 
		   \mathop{\longrightarrow}^{\langle \phi  \rangle} 
		          U(1)_Q \times SU(3)_C. 
\end{align}
At first, $SO(10)$ breaks down to the Pati-Salam group and D-parity invariance i.e, $SO(10) \rightarrow G_{224D}$ which can be achieved by giving a non-zero vev to $G_{224}$ singlet contained in a $\{54\}_H$-plet Higgs of $SO(10)$ (see refs.~\cite{Chang:1983fu, Chang:1984uy, Chang:1984qr} for detailed discussion on $SO(10)$ GUT breaking and D-parity invariance). This singlet scalar $\eta$ is even under D-parity. The subsequent stage of symmetry breaking $G_{224D} \rightarrow G_{224}$ is done with another singlet scalar odd under D-parity. The $G_{224}$ multiplet $\sigma \equiv (1,1,1) \subset \{210\}_H$, being odd under D-parity, is responsible for $G_{224D} \rightarrow G_{224}$ symmetry breaking. Here we denote this scale of D-parity invariance as $M_D$ ensuring equal value of $g_L$ and $g_R$.
The important stage of symmetry breaking $G_{224} \rightarrow G_\text{SM}$ is happened when right-handed Higgs field $\Omega_R (1,3,15)$ gets its non-zero vev. This symmetry breaking scale is fixed at few PeV scale such that decaying dark matter $\Sigma_{R}$, right-handed charged gauge boson $W_R$ and relevant scalars playing an important role in dark matter phenomenology and IceCube explanation -- all lie around that scale. The last stage of symmetry breaking $G_\text{SM} \rightarrow U(1)_{\rm em} \times SU(3)_C$ is done via SM Higgs doublet reproducing fermion masses and mixing. 
Here both manifest left-right theory with gauge group $SU(2)_L \times SU(2)_R \times U(1)_{B-L} \times SU(3)_C$ and Pati-Salam symmetry $SU(2)_L \times SU(2)_R \times SU(4)_C$ occur at same scale and hence, we only talk about Pati-Salam symmetry now onwards and its embedding in a non-SUSY $SO(10)$ GUT. The usual quarks and leptons are contained in Pati-Salam multiplet as $(2,1,4)_F + (2,1,\overline{4})_F\, \subset 16_F $. The fermion dark matter components transforming isospin triplet under $SU(2)_{L,R}$ are contained in Pati-Salam multiplet as $\Sigma_L(3,1,1)+\Sigma_R(1,3,1) \subset 45_F$. 
The complete spectrum of particles in the mass range $M_Z-M_R$, $M_R-M_{W_R}$, $M_{W_R}-M_D$ and $M_D-M_U$ is given in table \ref{tab4}. From phenomenological point of view, the left-handed fermion triplet kept at TeV scale and the corresponding right-handed fermion triplet is considered at Pati-Salam symmetry breaking scale i.e, at $10^{6}~$GeV in order to explain the PeV IceCube events. 
\begin{table}[h!]
\begin{center}
\begin{tabular}{|c|c|c|}
\hline
Group $G_{I}$  & Fermions      & Scalars       \\
\hline \hline
$\begin{array}{l}
G_{213} \\
(M_Z \leftrightarrow M_{R})  \end{array}$  & 
$\begin{array}{l}
Q_{L}(2,1/6,3) \\
u_{R}(1,2/3,3), d_{R}(1,-1/3,3) \\
\ell_{L}(2,-1/2,1), e_{R}(1,-1,1)
  \end{array}$ &
$\begin{array}{l}
\phi(2, \frac{1}{2},1) \end{array}$
 \\
\hline 
$\begin{array}{l}
{\small G_{2213}}\\
(M_{R} \leftrightarrow M_{W_R})  \end{array}$   &
${\small \begin{array}{l}
Q_L(2,1,1/3,3),Q_R(1,2,1/3,3) \\
\ell_L(2,1,-1,1),\ell_R(1,2,-1,1) \\ 
\Sigma_L(3,1,0,1), \Sigma_R(1,3,0,1)  \end{array}}$ &
${\small \begin{array}{l}
\Phi (2,2,0,1)   \\
\Delta_R(1,3,2,1), \Omega_R(1,3,0,1)   \\
\psi(3,3,0,1)  \end{array}}$                                                                   
\\
\hline 
$\begin{array}{l}
{\small G_{224}}\\
(M_{W_R} \leftrightarrow M_D)  \end{array}$   &
${\small \begin{array}{l}
\Psi_L(2,1,4), \Psi_R(1,2,\overline{4}) \\ 
\Sigma_L(3,1,1), \Sigma_R(1,3,1)  \end{array}}$ &
${\small \begin{array}{l}
\Phi(2,2,1), \\
\Delta_R(1,3,\overline{10}),\Omega_R(1,3,15) \\ 
\psi(3,3,1), \Sigma(1,1,15) \end{array}}$   
\\
\hline 
$\begin{array}{l}
{\small G_{224D}}\\
(M_D \leftrightarrow M_U)  \end{array}$   &
${\small \begin{array}{l}
\Psi_L(2,1,4), \Psi_R(1,2,\overline{4}) \\ 
\Sigma_L(3,1,1), \Sigma_R(1,3,1)  \end{array}}$ &
${\small \begin{array}{l}
\Phi(2,2,1) \\
\Delta_R(1,3,\overline{10}),\Delta_L(3,1,10) \\  
\Omega_R(1,3,15),\Omega_L(3,1,15) \\ 
\Sigma(1,1,15), \sigma(1,1,1) \\
\psi(3,3,1)\end{array}}$   
\\
\hline
\hline
\end{tabular}
\caption{Fermions and Scalars at different stages of symmetry breaking scales i.e, in the mass range $M_Z-M_R$, $M_R-M_{W_R}$, $M_{W_R}-M_D$ and 
$M_D-M_U$. Here $M_Z$ is the SM $Z$ boson mass, $M_R$ is the scale at which left-right symmetry breaks and $M_{W_R}$ is the scale at which 
Pati-Salam symmetry is broken down. In our numerical analysis, we consider both left-right symmetry breaking scale is very close to Pati-Salam symmetry breaking 
scale and thus, its effect in RG analysis is not accounted for simplicity. We also kept left-handed fermion triplet at few TeV scale while right-handed fermion triplet 
at few PeV scale. The one-loop beta coefficients derived in the mass range $M_{W_R}-M_D$ as $b_i=\{-2/3,33/3,-13/3 \}$ and for $M_{D}-M_U$ as $b_i=\{33/3,33/3,2/3 \}$ 
by taking $(3,3,1)$, $(3,1,15)$ and $(1,3,15)$ as real representations.}
\label{tab4}
\end{center}
\end{table}
\begin{figure}[!htbp]
\centering
\includegraphics[width=0.7\textwidth]{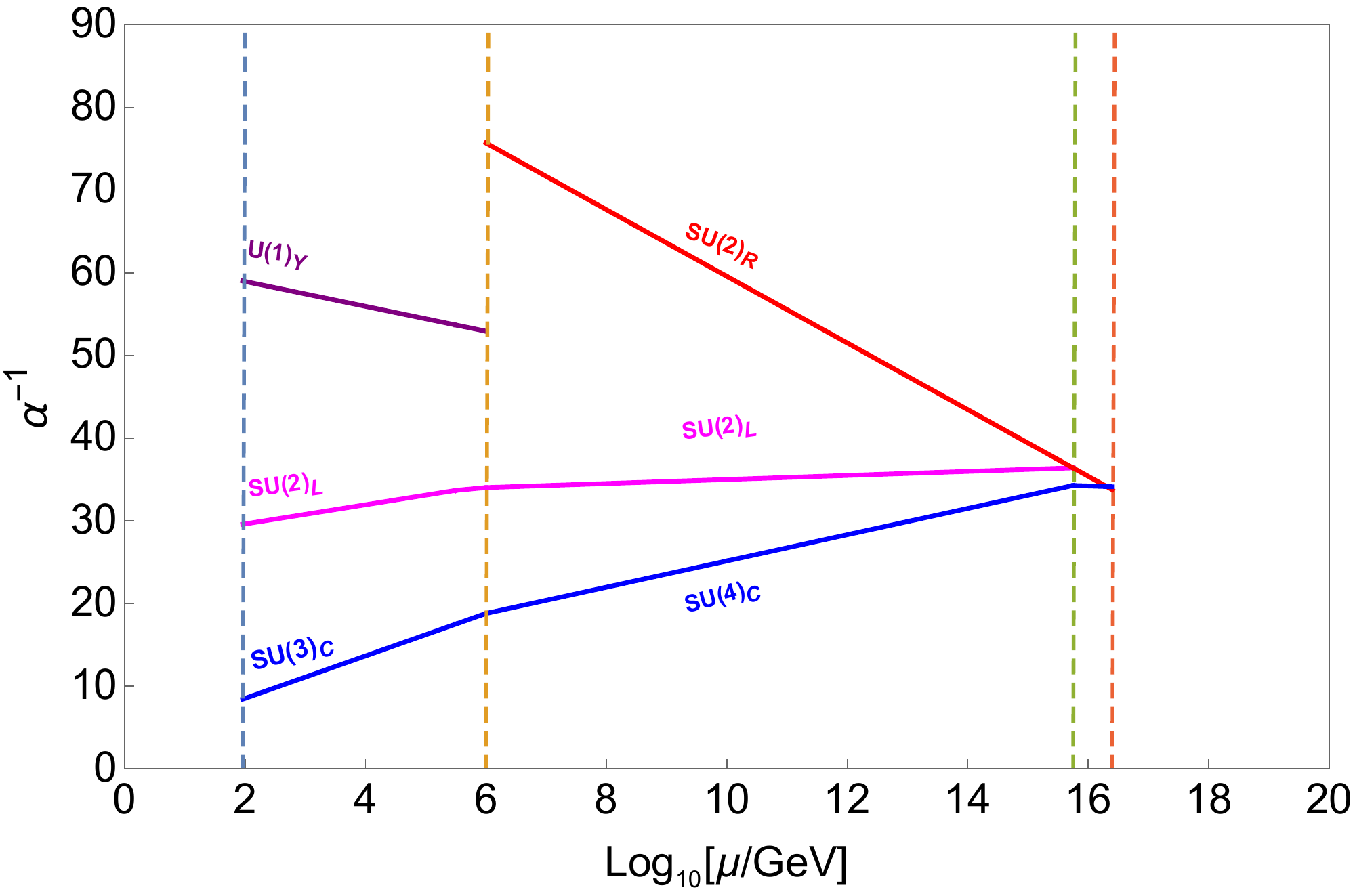}
\caption{Gauge coupling unification for a PeV scale left-right symmetric framework explaining IceCube data via decaying dark matter. The gauge couplings unify at $M_U=10^{16.85}~$GeV whereas other intermediate mass scales are $M_D \approx 10^{15.6}~$GeV (D-parity breaking scale), $M_R \approx 10^{6}~$GeV ($W_R$ mass at the Pati-Salam symmetry breaking scale).}
\label{fig6}
\end{figure}
The one-loop renormalisation group equations (RGEs) for gauge couplings $g_i$ for different stages of symmetry breaking scales is given by
\begin{equation}
\mu\,\frac{\partial g_{i}}{\partial \mu}=\frac{b_i}{16 \pi^2} g^{3}_{i},
\end{equation}
where the classic formula for one-loop beta-coefficients\footnote{The study of RG evolution of gauge couplings along with derivation for one-loop beta coefficients in a non-supersymmetric $SO(10)$ GUT with left-right symmetry and/or Pati-Salam intermediate symmetry breaking steps have been recently done in Refs~\cite{Patra:2014goa,Deppisch:2014qpa,Deppisch:2014zta,Hati:2017aez,Deppisch:2016scs,
Patra:2015bga,Awasthi:2013ff,Borah:2013lva,Bandyopadhyay:2017uwc,Arbelaez:2017ptu,Dev:2015pga}.} $b_i$ are given by
\begin{align}
	b_i = - \frac{11}{3} \mathcal{C}_{2}(G) 
			+ \frac{2}{3} \,\sum_{R_f} T(R_f) 
			  \prod_{j \neq i} d_j(R_f) 
            + \frac{1}{3} \sum_{R_s} T(R_s) 
              \prod_{j \neq i} d_j(R_s)\,.
\label{oneloop_bi}
\end{align}
The parameters used in the above formula for one-loop beta coefficients have their usual meaning as follows
\begin{itemize}
\item $\mathcal{C}_2(G)$: quadratic Casimir operator for gauge bosons which belong to their adjoint representation,
\begin{equation}
	\mathcal{C}_2(G) \equiv 
	\begin{cases} 
		N & \text{if } SU(N), \\
    0 & \text{if }  U(1).
	\end{cases}
\end{equation}
\item $T(R_f)$ and $T(R_s)$: traces of the irreducible representation for a fermion and scalar $R_{f,s}$, respectively,
\begin{equation}
	T(R_{f,s}) \equiv 
	\begin{cases} 
		1/2 & \text{if } R_{f,s} \text{ is fundamental}, \\
    N   & \text{if } R_{f,s} \text{ is adjoint}, \\
		0   & \text{if } U(1).
	\end{cases}
\end{equation}
\item $d(R_{f,s})$: dimension of a given fermion or scalar representation $R_{f,s}$ under non-abelian gauge group $SU(N)$ except the considered $i$-th~gauge group. 
\item Extra factor of $1/2$ should be multiplied if the Higgs belongs to a real representation. 
\end{itemize}
The RG evolution of gauge couplings is carried out for SM to Pati-Salam (or left-right symmetric theory) symmetry breaking scale, Pati-Salam scale with $g_L \neq g_R$ to Pati-Salam scale with D-Parity ($g_L=g_R$) and finally from Pati-Salam with D-Parity invariance scale to unification scale $M_U$. In figure~\ref{fig6}\,, we show that gauge couplings successfully unify at $10^{16.45}~$GeV with intermediate scales as $M_{R} \simeq 10^{6}~$GeV and $M_D=10^{15.6}~$GeV.

\section{Dark Matter}
\label{sec:relic}
In this model, we have two dark matter candidates namely the neutral components of fermion triplets $\Sigma_{L,R}$. We consider the right handed DM to be at PeV scale while the left handed one can be as light as a TeV. Usually, the relic density of Cold Dark Matter (CDM) in the Weakly Interacting Massive Particle (WIMP) scenario is calculated using the formalism discussed in ref. \cite{Jungman:1995df} which can be generalised to our present case as already discussed in \cite{Heeck:2015qra, Garcia-Cely:2015quu, Borah:2016ees} as
\begin{equation}
\Omega_{\rm DM} h^2 = \Omega_{\Sigma^0_L} h^2 + \Omega_{\Sigma^0_R} h^2.
\label{eq:relic}
\end{equation}
However, for generic thermal relic dark matter there exists an upper bound on the dark matter mass $(M_{\text{DM}} \leq \mathcal{O}(100) \; \text{TeV})$ coming from the unitarity limit on the annihilation cross section \cite{Griest:1989wd}. However, such a limit on dark matter mass can be relaxed if dark matter annihilation receives a Breit-Wigner enhancement \cite{Griest:1989wd, Ibe:2008ye}. As we will see below, the usual gauge interactions of the right handed fermion dark matter are not sufficient to give such an enhancement. Even if we include a separate scalar mediated annihilation channel, it is not sufficient the avoid the unitarity bound as we discuss below. 
\begin{figure}[h!]
\begin{center}
\includegraphics[width=0.9\textwidth]{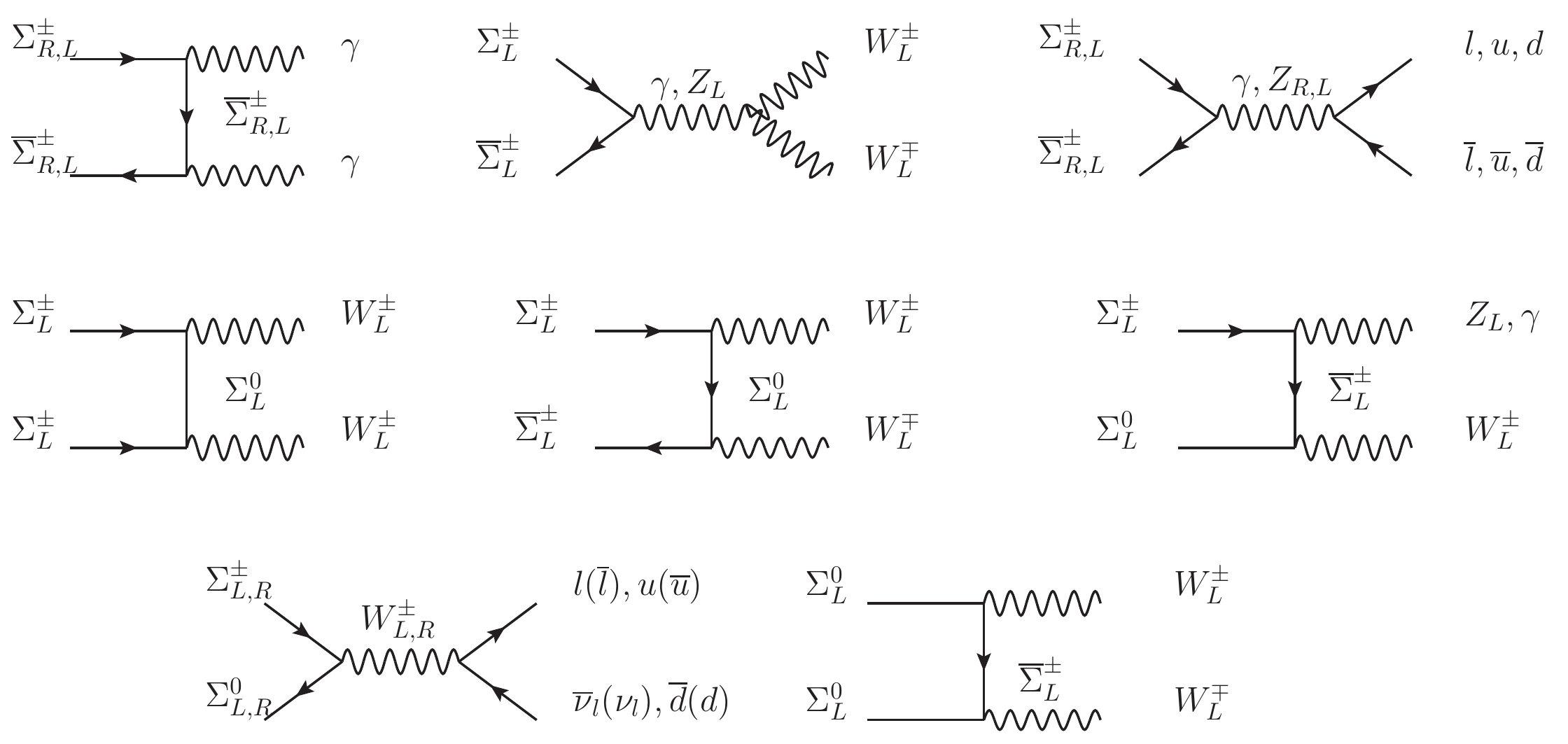}
\caption{The annihilation and coannihilation diagrams for fermion triplet dark matter relevant for relic density computation within the framework of minimal left-right dark matter ~\cite{Heeck:2015qra}.}
\label{fig1}
\end{center}
\end{figure}
The total abundance is constrained by the observed number mentioned in Eq.~\eqref{dm_relic} given by the observations from the Planck experiment~\cite{Ade:2015xua}. We can calculate the relic abundance of both the DM candidates independently in the limit of negligible left-right mixing of gauge bosons and negligible coannihilations between the two DM candidates mediated by the bitriplet scalar. The assumption of negligible coannihilation is justified in the limit of large mass splitting between the two DM candidates as we discuss below. In this decoupled limit, the individual relic abundance can be calculated by solving the corresponding Boltzmann equation
\begin{equation}
\frac{dn_{\chi}}{dt}+3Hn_{\chi} = -\langle \sigma v \rangle (n^2_{\chi} -(n^{\text{eqb}}_{\chi})^2)
\label{BE1}
\end{equation}
where $n_{\chi}$ is the dark matter number density and $n^{\text{eqb}}_{\chi}$ is the corresponding equilibrium number density. $H$ is the Hubble expansion rate of the Universe and $ \langle \sigma v \rangle $ is the thermally averaged annihilation cross section of the dark matter particle $\chi$. The approximate analytical solution of the above Boltzmann equation leads to the following expression for density parameter \cite{Kolb:1990vq, Scherrer:1985zt}
\begin{equation}
\Omega_{\chi} h^2 \approx \frac{1.04 \times 10^9 x_F}{M_{\text{Pl}} \sqrt{g_*} (a+3b/x_F)}
\end{equation}
where $x_F = m_{\chi}/T_F$, $m_{\chi}$ is the mass of dark matter particle, $T_F$ is the freeze-out temperature, $g_*$ is the number of relativistic degrees of freedom at the time of freeze-out and $M_{\text{Pl}} \approx 10^{19}$ GeV is the Planck mass. Here $a, b$ are the coefficients of partial wave expansion $ \langle \sigma v \rangle = a +b v^2$. The freeze-out temperature or $x_F$ can be calculated iteratively from the following relation 
\begin{equation}
x_F = \ln \left(\frac{0.038gM_{\text{Pl}}m_{\chi}\langle\sigma v \rangle}{g_*^{1/2}x_F^{1/2}}\right)
\label{xf}
\end{equation}
which follows from equating the interaction rate with the Hubble expansion rate of the Universe. The thermal averaged annihilation cross section $\langle \sigma v \rangle$ is given by \cite{Gondolo:1990dk}
\begin{equation}
\langle \sigma v \rangle = \frac{1}{8m^4_{\chi}T K^2_2(m_{\chi}/T)} \int^{\infty}_{4m^2_{\chi}} ds~\sigma (s-4m^2_{\chi})\sqrt{s}K_1(\sqrt{s}/T) 
\label{eq:sigmav}
\end{equation}
where $K_i$'s are modified Bessel functions of order $i$. Although we are assuming negligible coannihilations between the two DM candidates justified by their large mass splitting, there can be sizeable coannihilations within each DM multiplet, between the neutral and the charged components. At the classical level, the mass splitting between the charged and the neutral components of each DM multiplet is zero. However, a non-zero mass splitting can arise at one loop level given by \cite{Heeck:2015qra,Garcia-Cely:2015quu},
\begin{align}
M_{\Sigma_L^\pm}-M_{\Sigma_L^0} &\simeq  \alpha_2  M_{W} \sin^2 (\theta_W/2) + \mathcal{O}(M_{W}^3/M^2_\Sigma) \,,\\
\label{eq:RH_mass_splitting}
M_{\Sigma_R^\pm}-M_{\Sigma_R^0} &\simeq \frac{\alpha_2}{4\pi} \frac{g_R^2}{g_L^2} M \left[ f(r_{W_2}) - c_M^2 f(r_{Z_2})-s_W^2 s_M^2 f(r_{Z_1})- c_W^2 s_M^2 f(r_\gamma)\right],
\end{align}
where $s_M = \sin{\theta_M} \equiv \tan{\theta_W}, r_X = M_X/M$ and $$ f(r) \equiv 2 \int^1_0 dx (1+x) \log{[x^2+(1-x)x^2]} ~.$$ Here the one loop self-energy corrections through mediations of gauge bosons are presented within the square bracket of the second expression. For example,  the mass splitting with the approximation $M_\Sigma \gg M_{W_R}$ goes as $\alpha_{2} \left( M_{W_R} - c_M^2 M_{Z_R}\right)/2$. Due to such tiny one loop mass splitting, the next to lightest component of each DM multiplet can be thermally accessible during the dark matter freeze-out. In such a situation, the dark matter can coannihilate with the heavier components which then affects the relic abundance of dark matter. In the presence of coannihilations, the effective cross section is given by \cite{Griest:1990kh}
\begin{align}
\sigma_{\text{eff}} &= \sum_{i,j}^{N}\langle \sigma_{ij} v\rangle r_ir_j \nonumber \\
&= \sum_{i,j}^{N}\langle \sigma_{ij}v\rangle \frac{g_ig_j}{g^2_{\text{eff}}}(1+\Delta_i)^{3/2}(1+\Delta_j)^{3/2}\exp\big(-x_F(\Delta_i + \Delta_j)\big) \nonumber
\end{align}
where, $x_F = \frac{m_{\chi}}{T_F}$ and $\Delta_i = \frac{m_i-m_{\chi}}{m_{\chi}}$  and 
\begin{align}
g_{\text{eff}} &= \sum_{i=1}^{N}g_i(1+\Delta_i)^{3/2}\exp(-x_F\Delta_i).
\end{align}
The thermally averaged cross section can be written as
\begin{align}
\langle \sigma_{ij} v \rangle &= \frac{x_F}{8m^2_im^2_jm_{\chi}K_2((m_i/m_{\chi})x_F)K_2((m_j/m_{\chi})x_F)} \nonumber \\
& \qquad \times \int^{\infty}_{(m_i+m_j)^2}ds ~\sigma_{ij}(s-2(m_i^2+m_j^2)) \sqrt{s}K_1(\sqrt{s}x_F/m_{\chi}). \nonumber
\label{eq:thcs}
\end{align}
Since the effective degrees of freedom $g_{\text{eff}}$ decreases exponentially with increasing mass splitting between DM and heavier components, one can ignore such coannihilation effects for scenarios with large mass splitting. The effective annihilation cross-section for the fermion triplet can be written as 
\begin{align}
\langle \sigma v \rangle_{\Sigma_{A}} = \frac{g^2_{\Sigma^0_A}}{g^2_{\text{eff}}} \sigma(\Sigma^0_A \Sigma^0_A) 
        + 4 \frac{g_{\Sigma^0_A}  g_{\Sigma^\pm_A} }{g^2_{\text{eff}}} \sigma(\Sigma^0_A \Sigma^\pm_A) \left(1+\Delta_A \right)^{3/2} 
        \exp(-x \Delta_A) \notag \\
        + \frac{g^2_{\Sigma^\pm_A}}{g^2_{\text{eff}}} \left[ 2\sigma(\Sigma^\pm_A \Sigma^\pm_A)  
 +2 \sigma(\Sigma^+_A \Sigma^-_A) \right]\left(1+\Delta_A \right)^{2} \exp(-2x \Delta_A)
\end{align}
where $\Delta_A=(M_{\Sigma^\pm_A}-M_{\Sigma^0_A})/M_{\Sigma^0_A}$ is the mass splitting ratio and $x=M_{\Sigma^0_A}/T$. Here $A=L,R$ denotes the dark matter candidates.
\begin{figure}[t]
\begin{center}
\includegraphics[scale=0.65]{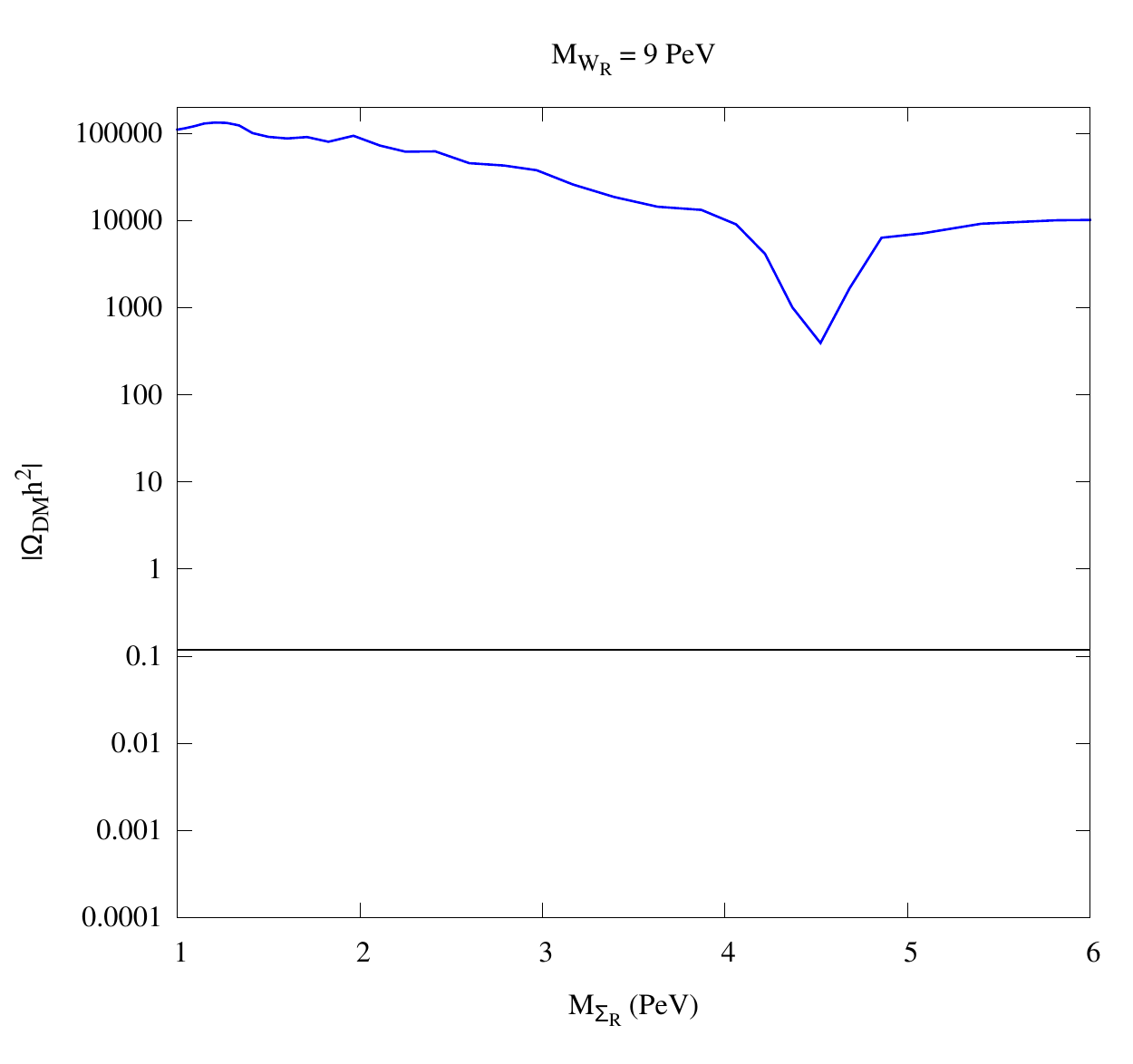}
\caption{Relic Abundance of heavy dark matter with only gauge annihilations. The black horizontal band correspond to $\Omega_{\text{DM}} h^2 = 0.1187 \pm 0.0017$.}
\label{fig31}
\end{center}
\end{figure}
\begin{figure}[t]
\begin{center}
\includegraphics[scale=0.65]{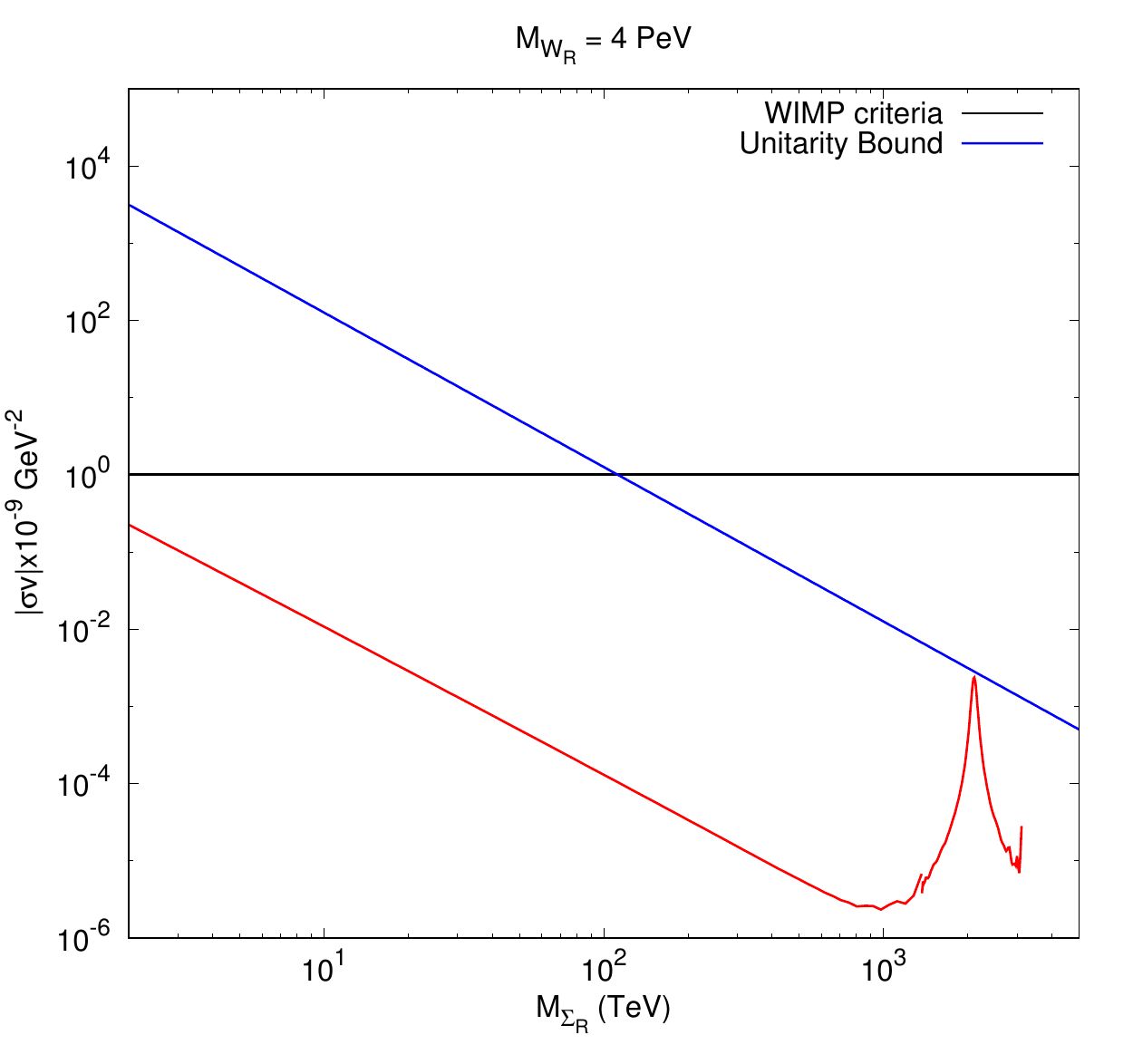}
\caption{Annihilation cross section of heavy dark matter (red line) in comparison to the cross section required to generate correct thermal relic abundance (black line) and the unitarity limit (blue line).}
\label{fig31a}
\end{center}
\end{figure}

Considering the heavier dark matter $\Sigma^0_R$ to be in the PeV regime we calculate its relic abundance by considering only gauge interactions. The gauge boson mediated annihilation and coannihilation channels are shown in figure \ref{fig1}. However, it was found for PeV right handed fermion dark matter that the gauge boson mediated channels were not sufficient to provide the resonance enhancement necessary to produce correct relic abundance of such superheavy dark matter candidates, as clearly seen from figure \ref{fig31}.  We then incorporate the presence of scalar triplet $\Omega_R$ through which the dark matter can (co)annihilate into a pair of light scalars from the bidoublet $\Phi$. One such possible process is the coannihilation of $\Sigma^0_R, \Sigma^{\pm}_R$ into $\Phi^0, \Phi^{\pm}$ with s-channel mediation of $\Omega^{\pm}_R$. Taking degenerate masses of $\Sigma^0_R, \Sigma^{\pm}_R$ as well as $\Phi^0, \Phi^{\pm}$ for illustrative purposes , the corresponding annihilation cross section is given by 
\begin{equation}
(\sigma v)_{\Sigma^0_R \Sigma^{\pm}_R \rightarrow \Phi^0, \Phi^{\pm}} = \frac{v^2_{S} f^2_S \lambda^2}{64 \pi M_{\Sigma^0_R}} \frac{(M^2_{\Sigma^0_R}-M^2_h)^{1/2} v^2}{(4 M^2_{\Sigma^0_R}-m^2_{\Omega^{\pm}_R})^2 + m^2_{\Omega^{\pm}_R} \Gamma^2_{\Omega^{\pm}_R}}
\end{equation}
where $\lambda, f_S$ are dimensionless couplings of $\Omega_R$ with $\Sigma_R$ and scalar bidoublet $\Phi$ respectively. As seen from the scalar potential in Eq.~\eqref{phiomega}, $f_S$ can be identified with $f_{15}$. $v_S$ is the vev of the neutral component of $\Omega_R$ which is taken to be similar to $v_R$. $m_{\Omega^{\pm}_R}, M_h$ are the masses of $\Omega^{\pm}_R, \Phi$ respectively. Since $\Omega_R$ does not couple directly to the SM fermions, it may be possible to tune its decay width $\Gamma_{\Omega^{\pm}_R}$ in such a way to give the required resonance enhancement to dark matter annihilations. This possibility was utilised in the previous work \cite{Dev:2016qbd} to bring the dark matter abundance into the observed range. However, as pointed out by the same authors later \cite{Dev:2016xcp, Dev:2016qeb}, one can not tune the decay width arbitrarily to enhance the cross section and hence the unitarity bound on the dark matter mass still prevails, in agreement with an earlier work \cite{Nussinov:2014qva}. This can be understood by approximating the above annihilation cross section for the resonance just above the threshold as
\begin{equation}
\sigma v \approx 16 \pi \frac{\Gamma_{\Sigma^0_R \Sigma^{\pm}_R} \Gamma_{\Phi^0, \Phi^{\pm}}}{(4 M^2_{\Sigma^0_R}-m^2_{\Omega^{\pm}_R})^2 + m^2_{\Omega^{\pm}_R} \Gamma^2_{\Omega^{\pm}_R}} \approx \frac{4\pi}{M^2_{\Sigma^0_R}}\text{BR}_{\Sigma^0_R \Sigma^{\pm}_R} \text{BR}_{\Phi^0, \Phi^{\pm}}
\end{equation}
where $\Gamma_{\Sigma^0_R \Sigma^{\pm}_R}, \Gamma_{\Phi^0, \Phi^{\pm}}$ are the partial decay widths of $\Omega^{\pm}_R$ into $\Sigma^0_R \Sigma^{\pm}_R$, $\Phi^0, \Phi^{\pm}$ respectively with $\text{BR}_{\Sigma^0_R \Sigma^{\pm}_R}, \text{BR}_{\Phi^0, \Phi^{\pm}}$ being the corresponding branching ratios. Therefore, even if we adjust the resonance, the annihilation cross section decreases with the mass of dark matter and hence it again leads to a unitarity bound $M_{\Sigma^0_R} \leq 100 \; \text{TeV}$ \cite{Griest:1989wd}. The authors of \cite{Nussinov:2014qva} found a slightly stronger bound $M_{\Sigma^0_R} \leq 20 \; \text{TeV}$ from the unitarity requirement. Thus for PeV scale heavy dark matter, we will overproduce it at least by a factor of around $10^6$, if the unitarity bound on the annihilation cross section is respected. This problem of producing the correct thermal relic abundance of PeV DM is also summarised in the plot shown in figure \ref{fig31a}. It shows that even near the resonance, the annihilation cross section remains much below the required one for producing the correct abundance. On the other hand, the required annihilation for correct relic abundance also lies above the maximum allowed cross section of DM from unitarity arguments.

The thermally overproduced PeV dark matter in our model can be reconciled with the observed dark matter relic abundance if there exists a long lived particle that decays into the standard model particles after the PeV dark matter freezes out. Such a decay should however occur before the big bang nucleosynthesis (BBN) temperature $T_{\text{BBN}} \sim \mathcal{O}$(MeV) in order to be consistent with successful BBN predictions. Such late decay of long lived particles can release extra entropy and dilute the abundance of PeV dark matter to bring it into the observed limit \cite{Scherrer:1984fd}. The dilution factor due to the decay of such a heavy long lived particle $N$ is given by \cite{Scherrer:1984fd}
\begin{equation}
\frac{1}{d} = \frac{s_{\text{after}}}{s_{\text{before}}} \approx 0.58 [g_*(T_r)]^{-1/4} \frac{\sqrt{\Gamma_N M_{\text{Pl}}}}{M_N Y_N}\;,
\end{equation}
where $\Gamma_N$ is the decay width of the heavy particle with mass $M_N$ and $Y_N=\frac{n}{s}$ is the initial abundance of the particle $N$ before it started to decay. Also, $g_*(T_r)$ is the relativistic degrees of freedom at a temperature $T_r$ just after the decay of $N$. This temperature to which the Universe cools down to following the release of entropy due to the decay of $N$ can be approximated as 
\begin{equation}
T_r \approx 0.78 [g_*(T_r)]^{-1/4} \sqrt{\Gamma_N M_{\text{Pl}}}\;.
\end{equation}
Although there can be several different possibilities of long lived particle in our model, we consider the late decay of $\Omega_R$ into the standard model particles.

\begin{figure}[t]
\begin{center}
\includegraphics[scale=0.75]{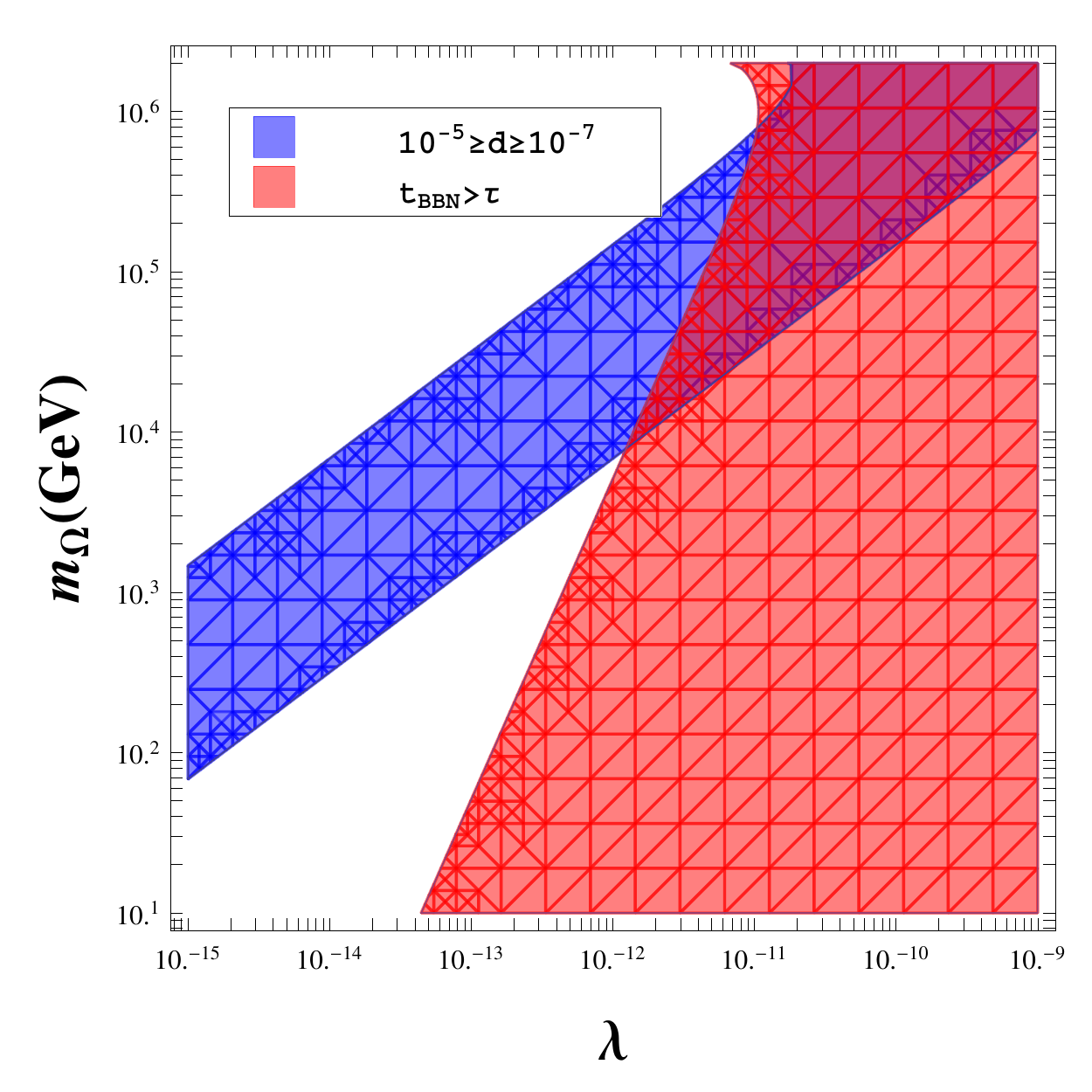}
\caption{Allowed parameter space from the requirement of producing correct dilution factor along with BBN constraints on lifetime.}
\label{entropy1}
\end{center}
\end{figure}

We can now find the decay diagrams through which the lightest component of $\Omega_R$ can decay. If the neutral component of $\Omega_R$  does not acquire any vev, and $M_{\Omega_R} < M_{\Sigma^0_R}$, then the lightest component of $\Omega_R$ can decay only at one-loop into the standard model particles. We consider the lightest component of $\Omega_R$ to be the neutral particle which can decay into a pair of photons through $\Sigma^{\pm}_R$ in loop. The Decay of the neutral component of the triplet $\Omega_R$ to two photons are given as
\begin{align}
\Gamma_{\Omega^0_R \rightarrow \gamma \gamma} &= \frac{\lambda^2 e^4}{32\pi m_{\Omega}}|\mathcal{I}|^2, \;\; |\mathcal{I}| = \frac{M^2_{\Sigma}}{256\pi^4}\left(4|\mathcal{A}|^2 + \Re(\mathcal{A}^*\mathcal{B})\right), \\
\mathcal{I}^{\mu\nu} &= \frac{iM_{\Sigma}}{16\pi^2}\left(\mathcal{A}g^{\mu \nu} + \frac{k^\mu_2 k^\nu_1}{m^2_{\Omega}}\mathcal{B}\right), \;\; \mathcal{A} = 1 + \frac{2t-1}{4}\ln \left[\frac{2t - 1 + \sqrt{1-4t}}{2t}\right]^2, \\
\mathcal{B} &= -\left(10 + 4\sqrt{1-4t}\ln \left[\frac{2t - 1 + \sqrt{1-4t}}{2t}\right] -2t\ln \left[\frac{2t -1 + \sqrt{1-4t}}{2t}\right]\right)
\end{align}
where $t = M^2_\Sigma/m^2_{\Omega}$ and $\lambda$ is the dimensionless couplings of $\Omega_R$ with $\Sigma_R$ appearing in Eq.~\eqref{sigmaomega}. We consider the initial abundance of the decaying particle 
$$Y_{\Omega_R} = \frac{n_{\Omega_R}}{s} = \frac{45}{\pi^4} \frac{\zeta(3)}{g_*(T_{f, \Omega_R})},$$
using $n_{\Omega_R} = \frac{2\zeta(3)}{\pi^2} T^3, s = \frac{2\pi^2}{45} g_*(T_{f, \Omega_R}) T^3$. Here $T_{f, \Omega_R}$ is the freeze-out temperature of $\Omega_R$ which is assumed to take place much before it starts decaying so that between freeze-out and decay, the $\Omega_R$ particles are neither being created nor destroyed. If we assume the freeze-out to be above the electroweak scale, then $g_*(T_{f, \Omega_R}) =106.75$ corresponding to all the SM degrees of freedom. It should be noted that here we are assuming equality between $g_{*s}$ appearing in entropy density expression and $g_*$ here which is valid at high temperatures where all species are in equilibrium with each other. We also take $g_*(T_r) = 10.75$, assuming that the decay is taking place very late, towards the end of QCD phase transition. This also automatically satisfies the criteria that the decay should take place after PeV dark matter freezes out. We now demand the dilution factor to be around $10^{-7}-10^{-5}$, the decay of the heavy particle to take place before BBN that corresponds to approximately 1 second after the big bang. These criteria constrain the parameter space in terms of $\Omega_R$ mass and its coupling to dark matter, i.e. $\lambda$, as seen from the plot shown in figure \ref{entropy1}. The upper limit of $\Omega_R$ mass comes from the requirement of forbidding tree level decay into $\Sigma^{\pm}_R$ pairs. It can be seen from the plot in figure \ref{entropy1} that the requirement of producing such a large entropy at late epochs, requires very long lifetime or equivalently, small decay width of the decaying particle forcing the coupling $\lambda$ to be very small in spite of one-loop suppression. Such fine-tunings can be relaxed to some extent if we consider $\Omega^{\pm}_R$ as the lightest component of $\Omega_R$ which can decay into a photon and an off-shell $W^{\pm}_R$. \\

\noindent
\textit{Naturalness of parameters:} In the above calculation, we have seen that the coupling between fermion dark matter candidates $\Sigma_{L,R}$ and the scalar triplets $\Omega_{L,R}$ has to be very small in order to have the correct entropy dilution mechanism at work or equivalently to make sure that $\Omega_R$ is long lived enough to release the required entropy during late epochs. Apart from this, one also requires additional fine tunings in order to make the heavier dark matter long lived and its decay into leptons plus the lighter dark matter more dominant over other decay channels. The additional scalar multiplets $\Omega_{L,R}, \psi$ do not couple to the SM fermions, but they can still decay fast into the SM Higgs by virtue of scalar couplings shown in Eqs.~\eqref{phiomega}, \eqref{deltaomega}, \eqref{psiomega} or into a pair of gauge bosons, if their neutral components acquire non-zero vev's. Therefore, the bitriplet scalar which mediates the heavier DM decay into lighter one plus leptons (as discussed below), can also open up decay modes into other standard model particles. All such decays can be forbidden by considering an approximate $U(1)_X$ global symmetry under which the particles shown in table \ref{tabdark} transform non-trivially. Such a symmetry prevents the fast decay of $\Omega^0_R, \Sigma^0_R$, both of which we have considered to be long-lived in this work. \begin{table}
\begin{center}
\begin{tabular}{|c|c|}
\hline
Particles & $U(1)_{X}$   \\
\hline
$\Sigma_L$ & $n_X$ \\
$\Sigma_R$ & $n_X$ \\
$\Omega_L$ & $n_X$ \\
$\Omega_R$ & $n_X$ \\
$\psi$ & $n_X$ \\
\hline
\end{tabular}
\end{center}
\caption{Fields having non-trivial transformation under $U(1)_X$ global symmetry}
\label{tabdark}
\end{table}
Clearly, such an approximate global symmetry can naturally suppress the Yukawa couplings like $f_{\psi}, \lambda$ appearing in equations \eqref{psisigma}, \eqref{sigmaomega} respectively as these trilinear terms violate the $U(1)_X$ global symmetry explicitly. Similarly, the trilinear mass term $\mu_{16}$ appearing in equation \eqref{psiomega} can also be naturally small, preventing the fast decay of $\Sigma^0_R \rightarrow \Sigma^0_L \Phi \Phi$.

One can also have decays like $\Sigma^0_R \rightarrow \Sigma^0_L W^+_L W^-_L$ or $\Sigma^0_R \rightarrow \Sigma^0_L Z_L Z_L$ if the neutral components of the scalar bitriplet acquire non-zero vev's. However, one needs to keep the vev of the bitriplet scalar small as the constraints on the electroweak $\rho$ parameter restricts it to $v_{\psi} \leq 2$ GeV \cite{Agashe:2014kda}. In the standard model, the $\rho$ parameter is unity at tree level, given by 
$$ \rho = \frac{M^2_{W_L}}{M^2_{Z_L} \cos^2 \theta_W} $$
where $\theta_W$ is the Weinberg angle. Any contribution to the masses of electroweak gauge bosons from scalar multiplets apart from the scalar doublet will give rise to deviation from the above tree level expression and hence will be constrained from electroweak precision measurements. The tiny vev of the neutral components of $\psi$ can be guaranteed if $\psi$ also transforms non-trivially under the approximate global symmetry $U(1)_X$, as shown in the table \ref{tabdark}. Such symmetry will naturally explain the vanishingly small trilinear mass term $\mu_{16}$ in Eq.~\eqref{psiomega} which can automatically guarantee a tiny vev to the neutral component of bitriplet scalar $\psi$. From the minimisation of the scalar potential with respect to $\psi$, it is straightforward to find the induced vev to be 
$$ \langle \psi \rangle \approx - \frac{\mu_{16} \langle \Phi \rangle^2}{\mu^2_{\psi}} $$
where $\mu^2_{\psi} >0 $ is the bare mass squared term of the bitriplet scalar. Therefore, a tiny mass term $\mu_{16}$ will naturally ensure the smallness of the vev. Such an approximate symmetry will also give rise to a small trilinear term $\mu_{236}$ Eq.~\eqref{psiomega} which will appear in the decay of heavier DM to lighter DM as we discuss in the next section. Therefore, an approximate global $U(1)_X$ symmetry can suppress the fast decay of heavier DM into the lighter one and a pair of $\Phi$ or $W^{\pm}_L$ or $Z_L$ bosons.

Similarly, one can also have fast decay of $\Omega^0_R$ into the standard model fields if it acquires a non-zero vev. In the discussion on thermal abundance of heavier dark matter, we considered non-zero vev of $\Omega^0_R$ in order to allow the s-channel coannihilation into light scalars. However, since that was not sufficient to produce the correct thermal abundance, we abandon that setup and have moved on to discussing the decay of $\Omega^0_R$ as mentioned above. In fact, a non-zero vev of $\Omega^0_R$ will allow it to decay very fast into the SM fermions mediated by $W_R$ bosons. Even if we assume a zero vev of $\Omega^0_R$ by assuming its bare mass squared term to be positive definite, it can still acquire a non-zero induced vev by virtue of its coupling with the scalar bidoublet shown in Eq.~\eqref{phiomega}. This induced vev can be found from the minimisation of the scalar potential as 
$$ \langle \Omega^0_R \rangle \approx - \frac{\mu_{15} \langle \Phi \rangle^2}{\mu^2_{\Omega}} $$
where $\mu^2_{\Omega} >0 $ is the bare mass squared term of the triplet scalar $\Omega_R$. The same trilinear interaction involving $\mu_{15}$ also induces the decay of $\Omega^0_R$ into a pair of SM Higgs bosons. In order to prevent $\Omega^0_R$ from acquiring a non-zero vev and from decaying into a pair of SM Higgs bosons we need to prevent its trilinear interactions with the bidoublet or make those interactions very small. This is possible by virtue of the non-trivial charge of $\Omega_{L,R}$ under the approximate global symmetry, as shown in table \ref{tabdark}. In particular, the trilinear mass terms $\mu_{14}, \mu_{15}$ in Eq.~\eqref{phiomega} can be vanishingly small as they explicitly break the $U(1)_X$ global symmetry, in accordance with the naturalness criteria.

Therefore, the presence of an approximate global symmetry $U(1)_X$ can naturally give rise to long lived heavier DM, long lived $\Omega^0_R$ required to release entropy at late epochs. Although such an approximate symmetry can naturally explain the long life of these two particles, it however can not explain the dominance of leptonic final state decay mode of heavier dark matter over other decay modes with non-leptonic final states like a pair of $\Phi$ or $W^{\pm}_L$ or $Z_L$ bosons in the final state. Since such approximate global symmetries are likely to originate from an ultraviolet complete theory, we leave this question to the details of such a theory at high energy scale.
\begin{figure}[t!]
\begin{center}
\includegraphics[width=0.6\textwidth]{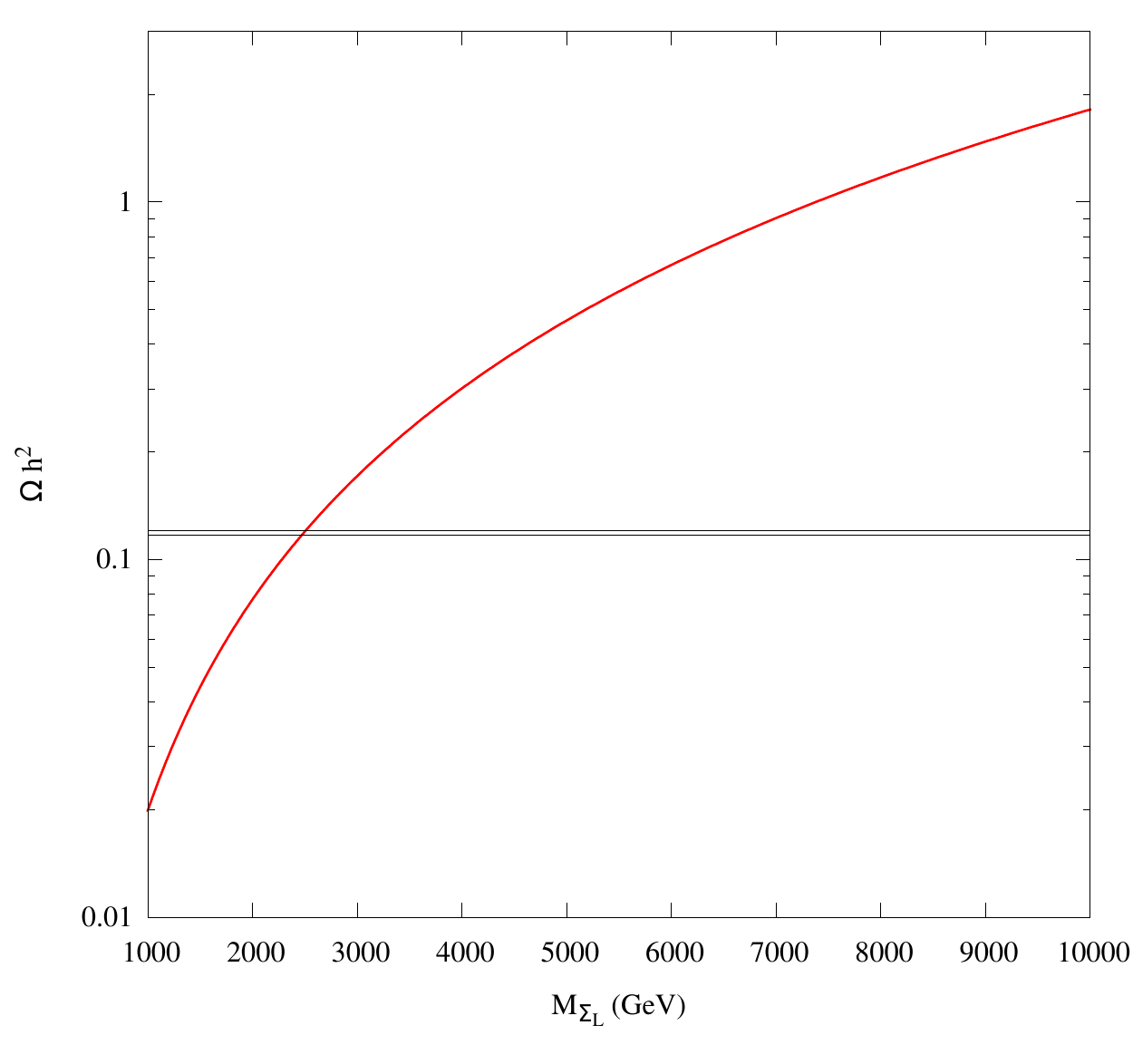}
\caption{The relic abundance of left fermion triplet dark matter as a function of its mass. The black horizontal band correspond to $\Omega_{\text{DM}} h^2 = 0.1187 \pm 0.0017$.}
\label{figLHDM}
\end{center}
\end{figure}
\begin{figure}[t]
\begin{center}
\includegraphics[scale=0.65]{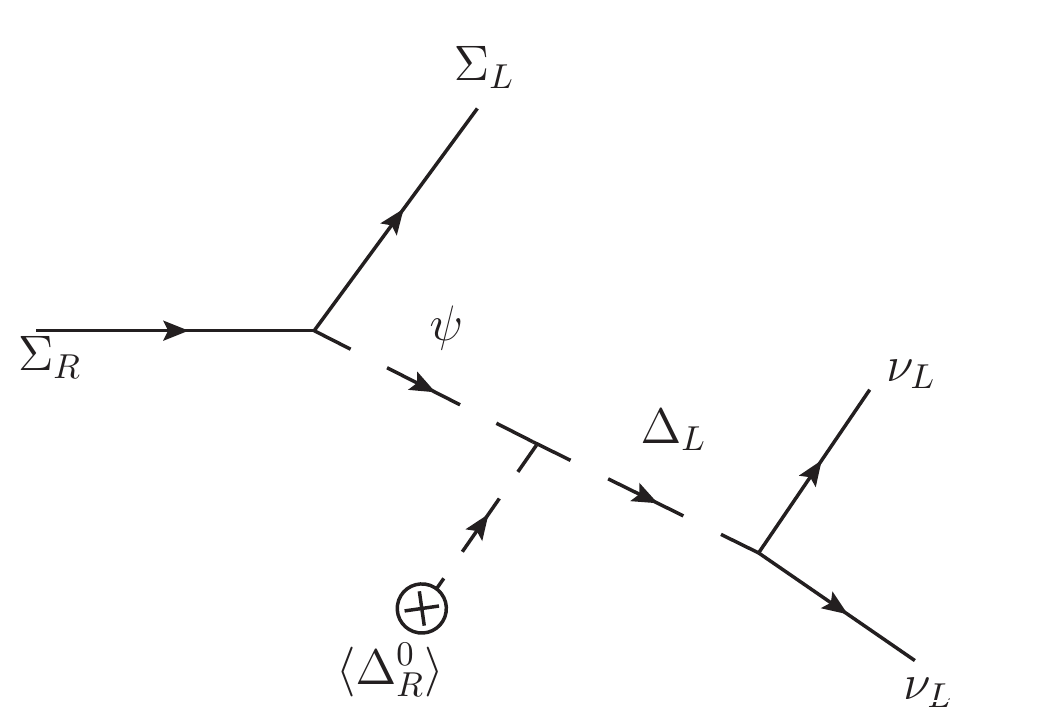}
\caption{Feynman diagrams for next to lightest dark matter particle decaying into lightest dark matter particle and two light neutrinos.}
\label{fig3}
\end{center}
\end{figure}

%

%
We also calculate the relic abundance of left fermion triplet dark matter. Since the mass splitting between different components of the left fermion triplet is a function of electroweak gauge boson mass, the only free parameter in this case is the mass of dark matter. We show the variation of its relic abundance with mass in figure \ref{figLHDM}. It can be seen from the figure that correct relic abundance can be generated for left fermion triplet mass around 2.5 TeV. For heavier masses, the relic is overproduced whereas it is under-abundant for lower masses. Since the left fermion triplet is the lighter dark matter candidate in our model and hence cosmologically stable, it is important to make sure that it does not get overproduced. In our subsequent analysis, we will assume the heavier dark matter to be the most dominant component which restricts the left triplet dark matter masses to be smaller than 2.5 TeV. Such TeV-scale mass for left fermion triplet is also taken into account in the RGE analysis discussed in the previous section.

\begin{figure}[t]
\centering
\includegraphics[scale=0.85]{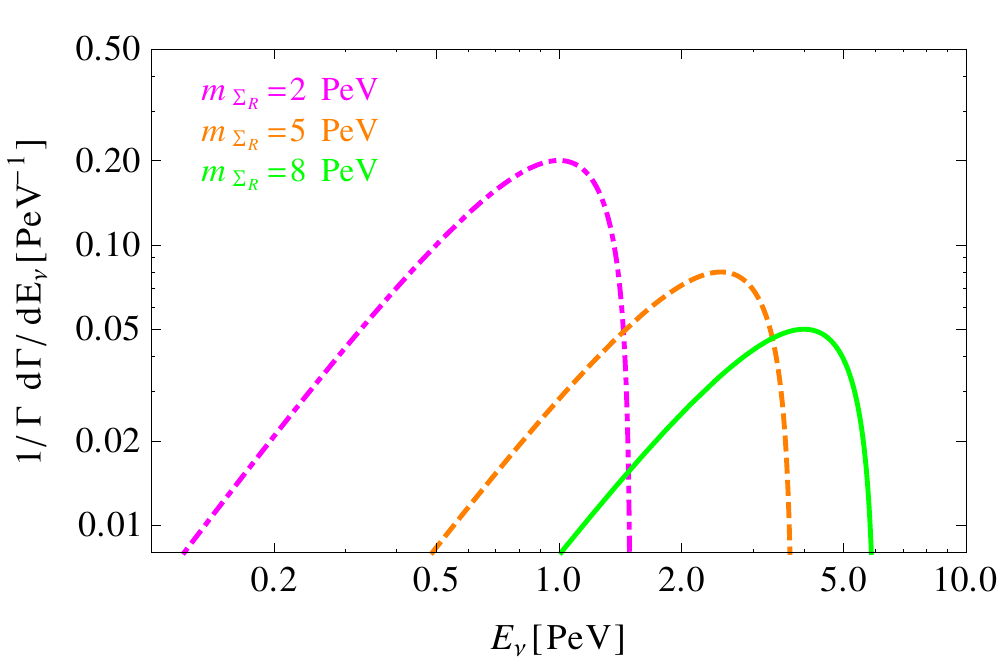}
\caption{Normalised energy distribution of the primary neutrinos coming from the DM decay $\Sigma_{R}^{0}\to \Sigma_{L}^{0}\bar{\nu}_{\ell}\nu_{\ell}$ for various masses of $\Sigma_{R}^{0}$.}
\label{fig:EnuDistr}
\end{figure}

Apart from the total relic abundance criteria to be satisfied by the two dark matter candidates together, the heavier dark matter decay should be consistent with the observed neutrino flux at IceCube experiment. The amplitude of the heavier dark matter decay diagram \ref{fig3} is given as,
\begin{align}
|\mathcal{M}|^2 &=2|f_{\psi} f_L \theta |^2\frac{z_1(1-z_1)}{(1-z_1-r^2_S)^2},
\end{align}
where $z_1 = 2p.p1/M^2_{\Sigma^0_R}\equiv 2E_{\Sigma_L}/M_{\Sigma^0_R}$, $r_S = M_S/M_{\Sigma^0_R}$ with $f_{\psi}, f_L$ being Yukawa couplings appearing in Eqs.~\eqref{psisigma}, \eqref{treeY} respectively, 
$\theta$ being the scalar mixing parameter, and $M_S$ being the mass of the neutral physical scalar as mediator. The mixing between the scalars $\psi, \Delta_L$ can be given by the angle $\theta$ as
$$\tan{2 \theta} = \frac{\mu_{236} v_R}{m^2_{\psi}-m^2_{\Delta_L}} $$
where $\mu_{236}$ is the trilinear mass term appearing in Eq.~\eqref{psiomega}. After doing further calculation, 
we get the following differential decay width,
\begin{align}
 \frac{\partial \Gamma}{\partial z_2} &= \frac{|f_{\psi} f_L \theta |^2 M_{\Sigma^0_R} }{256\pi^3}
 \left(-1+r^2_S-z_2 +\frac{r^2_S(1-r^2_S)}{r^2_S-z_2}+(1-2 r^2_S)\mbox{ln}\left[\frac{r^2_S-z_2}{r^2_S}\right]\right),
\end{align}
where $z_2= 2E_\nu/M_{\Sigma^0_R}$. Integrating the differential decay width over the neutrino energy gives the following decay width of dark matter, 
\begin{align}
 \Gamma &=  \frac{|f_{\psi} f_L \theta |^2 M_{\Sigma^0_R}}{256\pi^3}\left[-\frac{5}{2} +3r^2_S + (1-4r^2_S + 3r^4_S)\ln \left[\frac{r^2_S-1}{r^2_S}\right]\right].
\end{align}
Assuming $M_S \gg M_{\Sigma^0_R}$, we get the neutrino energy distribution (for massless final states),
\begin{equation}
 \frac{1}{\Gamma}\frac{\partial \Gamma}{\partial E_\nu} = \frac{24}{5}\frac{E^2_\nu}{ M^3_{\Sigma^0_R}}\left(1-\frac{4}{3}\frac{E_\nu}{M_{\Sigma^0_R}}\right).
\end{equation}
The normalised energy distribution of the primary neutrinos is shown in figure~\ref{fig:EnuDistr} for three different values of heavier dark matter masses. From figure~\ref{fig:EnuDistr} it is clear that PeV events at IceCube can be described by DM decay with its mass of about 5 PeV.

\section{IceCube PeV Events}
\label{sec:ic}
The 1347-day IceCube high energy starting events (HESE) neutrino data shows 54 events with deposited energies ranging between 30 TeV to 2.1 PeV~\cite{Aartsen:2015zva, Aartsen:2013jdh, Vincent:2016nut}. That the origin of the events above a few tens of TeV is non-atmospheric and extra-terrestrial in nature has been established with a high level of significance~\cite{Aartsen:2013jdh}. The three highest energy events of deposited energies 1.04 PeV, 1.14 PeV and 2.0 PeV are all shower events~\cite{Aartsen:2014gkd}. In addition, recently a track-like event with deposited energy $\sim 3$ PeV is reported by IceCube~\cite{Aartsen2016}. Its explanation as a signal from a new astrophysial neutrino flux is explored in~\cite{Kistler2016}, suggesting a window into astrophysical neutrinos at $E_\nu \sim 100$ PeV. Moreover, the highly sought after Glashow resonance, with the energy around 6.3 PeV, has not yet been observed\footnote{The absence of the Glashow resonance is discussed in details in refs.\cite{Anchordoqui2014, Learned2014, Tomar2015}}. A single power-law fit to the IceCube flux in the expected $E^{-2}$ spectrum, coming from Fermi shock acceleration considerations, predicts more number of multi-PeV events, it is thus disfavoured. 
It has been shown that the scenarios with superheavy decaying dark matter can ameliorate some of the previously mentioned tensions~\cite{Esmaili:2013gha, Feldstein:2013kka, Ema:2013nda, Bhattacharya:2014vwa, Rott:2014kfa, Esmaili:2014rma, Bhattacharya:2014yha, Cherry:2014xra, Fong:2014bsa, Murase:2015gea, Boucenna:2015tra, Ko:2015nma, Dev:2016qbd, Bhattacharya:2016tma}. In the present model that we are considering we have a two-components ($\Sigma_{L}^{0}$ and $\Sigma_{R}^{0}$) DM scenario. The mass of  the heavier one among these two species, the $\Sigma_{R}^{0}$ is in PeV range. We want to test its feasibility as a possible candidate which can explain the IceCube PeV events through its three-body decay as $\Sigma_{R}^{0} \to \Sigma_{L}^{0} \overline{\nu_\ell}\nu_\ell$, where $\ell \equiv e,\mu,\tau$. Clearly, the primary production of neutrinos will contribute in the incoming neutrino flux. 
The other decay channels of $\Sigma^0_{R}$ includes $\Sigma^0_{R} \to \Sigma^0_{L}\overline{\ell} \ell$, where $\ell \equiv e,\mu,\tau$. It is clear that in this model, the secondary production of neutrinos is possible through the decays of primary leptons ($\mu$ and $\tau$) which contributes in the lower energy range of the event distribution at IceCube. It is important to mention that the astrophysical contribution provides a good fit to the low-energy data and any additional contribution is tightly constrained from IceCube data~\cite{Aartsen:2015zva}. In our considered model, by suppressing the decay of $\Sigma^0_{R}$ into quarks and bosons, it is possible to avoid these constraints.
The neutrino flux coming from the heavy DM decay has two components, namely the galactic and extra-galactic components. In the subsequent, we briefly elaborate the standard method to calculate this flux following the procedure prescribed in~\cite{Esmaili:2012us, Esmaili:2013gha, Esmaili:2014rma}. 
The galactic component comes from the decay of DM in the Milky Way halo and the differential flux corresponding to this is given by,
\begin{align}
\frac{d\Phi^{\rm G}}{dE_{\nu}} = 
            \frac{1}
            {4\pi m_{\Sigma_{R}^{0}} \tau_{\Sigma_{R}^{0}}}
            \frac{dN(E_{\nu})}{dE_{\nu}}
            \int_{0}^{\infty} ds ~ 
            \rho_{\Sigma_{R}^{0}}(r(s,l,b)),
\end{align}
where $m_{\Sigma_{R}^{0}}$ and $\tau_{\Sigma_{R}^{0}}$ denote the mass and lifetime of the DM particle $\Sigma_{R}^{0}$ and $dN(E_{\nu})/dE_{\nu}$ is the energy spectrum of neutrinos produced in the decay of the DM. We compute $dN(E_{\nu})/dE_{\nu}$ for primary as well as secondary neutrinos using the methods outlined in ref.~\cite{Cirelli2011, Cline2013}. For the considered spectrum we implicitly assumed the sum over all neutrino and anti-neutrino flavors. The quantity $\rho_{\Sigma_{R}^{0}}(r)$ is the density profile of DM particles in our galaxy as a function of distance from the galactic centre. In our calculations we take the density profile to be of the NFW form~\cite{Navarro:1996gj} $\rho_{\Sigma_{R}^{0}}(r) = \rho_{0}(r_{0}/r)/(1 + r/r_{0})^{2}$ with $\rho_{0} = 0.33$ GeV cm$^{-3}$ and $r_{0} = 20$ kpc. The integral over the parameter $s$ is basically the line-of-sight integral to obtain the flux at the Earth and $s$ is related to $r$ via the relation $r(s,l,b) = \sqrt{s^{2} + R_{\odot}^{2} - 2sR_{\odot}\cos b\cos l}$, where $R_{\odot} \simeq 8.5$ kpc is the distance of the Sun from the galactic centre; $l$ and $b$ represent the longitude and latitude in galactic coordinates.
The second component, i.e., the extra-galactic component comes from the DM decays at cosmological scales and this produces an isotropic diffuse flux which in its differential from is given by,
\begin{align}
\frac{d\Phi^{\rm EG}}{dE_{\nu}} = 
           \frac{\Omega_{\rm DM}\rho_{c}}
           {4\pi m_{\Sigma_{R}^{0}} \tau_{\Sigma_{R}^{0}}}
           \int_{0}^{\infty} dz \frac{1}{H(z)}
           \frac{dN((1+z)E_{\nu})}{dE_{\nu}}, 
\end{align}
where $\rho_{c} = 5.5\times 10^{-6}$ GeV cm$^{-3}$ is the critical density of the universe and the Hubble expansion rate as a function of the redshift $z$ is given by $H(z) = H_{0}\sqrt{\Omega_{\Lambda} + \Omega_{\rm m}(1+z)^{3}}$. For the considered parameters we use the results from Planck collaboration~\cite{Ade:2015xua} which, following the standard $\Lambda$CDM cosmology gives $\Omega_{\rm DM} = 0.27$, $\Omega_{\Lambda} = 0.68$, $\Omega_{\rm m} = 0.32$, and $h\equiv H_{0}/100$ km s$^{-1}$Mpc$^{-1}$ = 0.67.
The total neutrino flux coming from DM decay is, can be written as the sum of galactic and extra-galactic contributions and given as,
\begin{align}
\frac{d\Phi}{dE_{\nu}} = \frac{1}{4\pi}
            \int d\Omega \left(
            \frac{d\Phi^{\rm G}}{dE_{\nu}}
            + \frac{d\Phi^{\rm EG}}{dE_{\nu}} \right).
\end{align}
For completeness we also take into account the astrophysical neutrino flux, the contributions of which comes from extragalactic supernova remnants (SNR), hypernova remnants (HNR)~\cite{Chakraborty2015}, AGN~\cite{Stecker1991, Kalashev2015}, and GRB~\cite{Waxman1997}. In order to parametrize the astrophysical flux, we assume a single unbroken power-law astrophysical flux:
\begin{equation}
 E^2_\nu \frac{d\Phi^{\mbox{\tiny astro}}_\nu(E_\nu)}{dE_\nu}= \Phi_0 \left(\frac{E_\nu}{100~\mbox{TeV}}\right)^{-\gamma},
\end{equation}
where $\Phi_0=2.2\times 10^{-8}~\rm GeV cm^{-2} sec^{-1} sr^{-1}$ and $\gamma=0.58$ which correspond to the best-fit value at Icecube~\cite{Aartsen:2015zva} assuming equal composition of neutrino flavor on Earth\footnote{The flavor composition of neutrino on the Earth depends on their flavor ratio at the source~\cite{Barger2014}. So the contribution from the decay of DM can, in principle, alter the neutrino flavor composition on the Earth. A comprehensive study of shower/track composition of the events can shed light on the more precise flavor ratio. Owing to the low available statistics such a study is not feasible~\cite{Chen:2014gxa, Palomares-Ruiz:2015mka}. In our analysis we assume the canonical $1:1:1$ flavor ratio.}.
The total neutrino flux will thus be the sum of the astrophysical flux and the flux coming from DM decay and this can be written as,
\begin{align}
\frac{d\Phi^{\rm tot}}{dE_{\nu}} = 
            \frac{d\Phi}{dE_{\nu}}
            + \frac{d\Phi^{\mbox{\tiny astro}}_\nu(E_\nu)}{dE_\nu}\;.
\end{align}
\begin{figure}[t]
\centering
\includegraphics[width=0.6\textwidth]{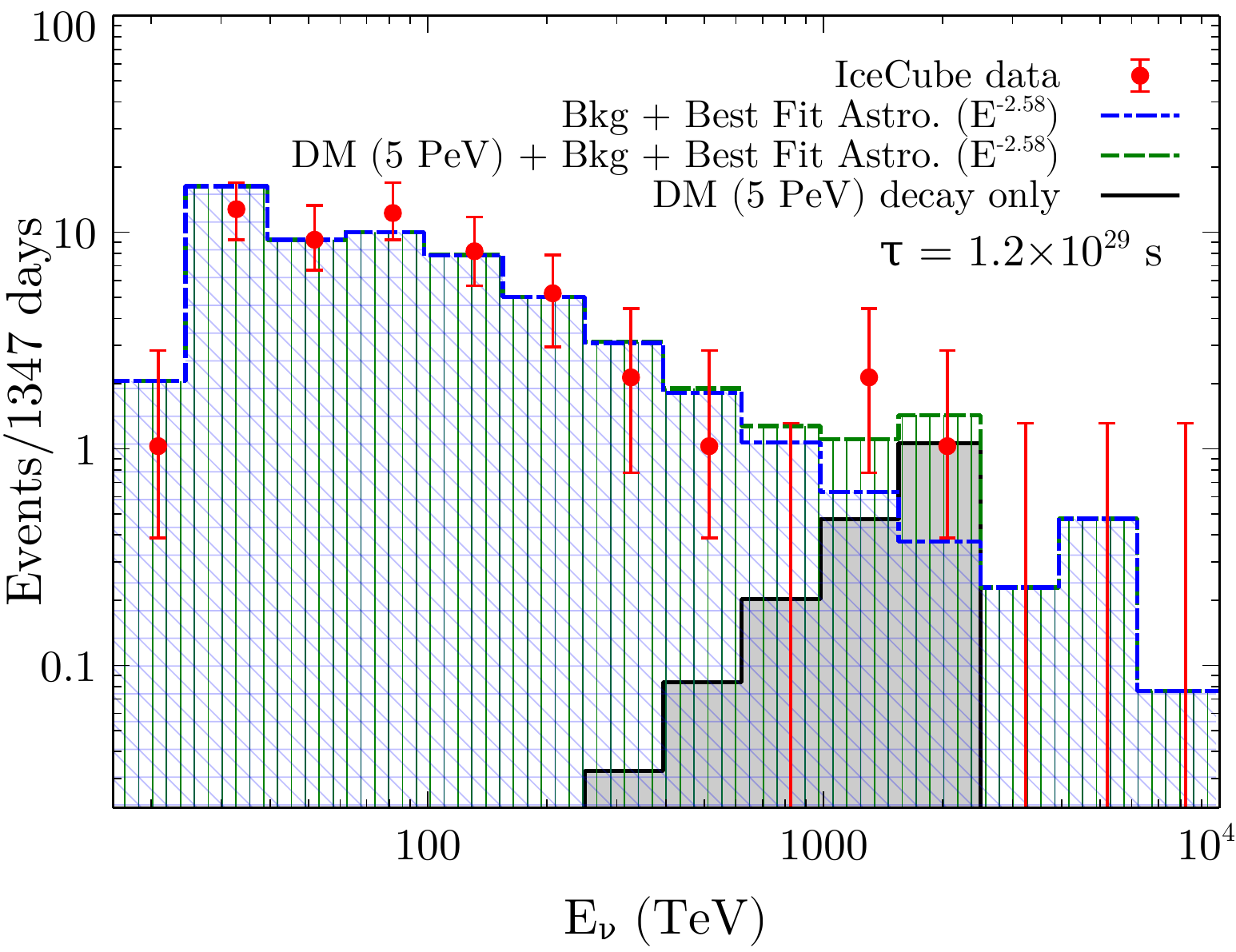}
\caption{Event distribution for 4-yr IceCube data with the contribution of decaying DM. Here we take mass of the DM to be 5 PeV with a lifetime of $1.2\times 10^{29}$sec.}
\label{fig:ICevt}
\end{figure}
After getting the total neutrino flux, it is now straightforward to compute the number of events and compare with the IceCube data. To this end we define the quantity $N_{i}$ which gives the number of events in a given deposited energy bin $[E_{i},E_{i+1}]$ as follows,
\begin{align}
N_{i} = t \int_{E_{i}}^{E_{i+1}}dE \frac{d\Phi^{\rm tot}(E_{\nu})}{dE_{\nu}} A(E_{\nu})\;,
\label{noe}
\end{align}
where $t$ is the exposure time which is 1347 days, $A(E_{\nu})$ is the neutrino effective area~\cite{Aartsen:2013jdh}. To compare the prediction of (\ref{noe}) with the IceCube observations, one needs to transmute the number of events in the terms of deposited energy. But as discussed in~\cite{Boucenna:2015tra} due to low statistics collected till now, it is acceptable to assume that the two energies coincide. It is important to mention that in the calculation of $N_{i}$ we have summed over all neutrino flavors.
For a characteristic benchmark point with $m_{\Sigma_{R}^{0}} = 5$ PeV and $\tau_{\Sigma_{R}^{0}} = 1.2\times 10^{29}$sec, we present the event distribution at the IceCube in figure~\ref{fig:ICevt}. From this figure it can be seen that the contribution in the neutrino flux coming from the decay of the DM, $\Sigma_{R}^{0}$ can explain the excess in the observed IceCube PeV neutrino events and the subsequent cut-off in the distribution. The lower energy region of the distribution is, however, in agreement with the astrophysical and atmospheric neutrino flux which constitutes the main background events and the contribution in the flux from the DM decay is negligible in this region.

\section{Summary and Conclusion}
\label{sec:concl}
Motivated by the possibility of dark matter interpretation of the IceCube high energy neutrino events we study a non-supersymmetric left-right model that not only explains the IceCube events from a heavy PeV dark matter decay, but also achieves gauge coupling unification at high energy scale with $SO(10)$ embedding. Although PeV scale heavy dark matter does not fall in the typical weakly interacting massive particle ballpark due to the unitarity bound on dark matter mass, our model can evade such bounds through resonance enhancement of dark matter annihilations. Also, the particle content of the model is chosen in such a way that predicts a PeV scale intermediate $SU(2)_R \times U(1)_{B-L}$ symmetry, suggesting a fundamental reason behind the origin of PeV scale dark matter, apart from its motivation from IceCube point of view.
The model has a pair of left and right fermion triplet dark matter candidates. Their equal masses in the limit of exact left-right symmetry acquire a large splitting due to the breaking of discrete left-right symmetry at a very high energy scale, resulting in a PeV scale heavy right handed fermion dark matter and a light TeV scale left handed fermion dark matter. The PeV scale dark matter can decay into the lighter one along with a pair of neutrinos on cosmological scale so that the high energy IceCube events can be explained. The large mass splitting between left and right fermion dark matter keeps their coannihilations sub-dominant. Also, in the limit of negligible left-right mixing, the annihilation of heavier dark matter into lighter ones can be ignored. This simplifies the dark matter relic abundance calculations as it can be calculated separately for individual dark matter candidates. The individual dark matter multiplets can however, have coannihilations within its components due to tiny mass splittings that arise at one loop level. We first calculate the heavier dark matter abundance by incorporating these coannihilations through right handed gauge bosons and show that these gauge interactions are not sufficient to give the required resonance enhancement to dark matter annihilations in order to bring the relic abundance down to the observed one. We then show that even if we introduce another scalar to give further s-channel enhancement, it is not possible to tune its decay width arbitrarily in order to generate the required annihilation cross section. The unitarity limit on the cross section forces us to restrict the dark matter mass to be at most 20 TeV. This overproduces the heavy dark matter by a factor of $10^5-10^6$ compared to the observed abundance. We then show one possible way of diluting this abundance by late decay of a heavy scalar which releases entropy through its decay before BBN and dilutes the thermally overproduced PeV dark matter by appropriate amount. Here we assume the heavier dark matter to be the dominant dark matter component to avoid any suppression in its decay flux into the high energy neutrinos. The lighter dark matter relic abundance can be easily kept at a suppressed value by choosing its mass suitably in the TeV regime. Unlike the PeV right handed dark matter, the TeV left handed dark matter annihilation can be sufficiently enhanced to keep the relic abundance suppressed even with electroweak scale gauge bosons as mediators.
The decay of PeV right handed dark matter to the TeV left handed dark matter with a pair of neutrinos can explain the high energy IceCube events. Using 4-yr IceCube data set of 54 events we show that together with astrophysical neutrino flux for lower energy events and the contribution in neutrino flux coming from the decay of the PeV right handed dark matter can fit the IceCube data from TeV-PeV energy range. We illustrate the fitting with a benchmark mass of the 5 PeV and lifetime of $\sim 10^{29}$ s. Clearly, future data with more statistics and precision will put more restrictions on this model. In passing it is worth mentioning that our benchmark evades the recent strong constraints that normally come from the gamma ray observations. In the present scenario, PeV right handed dark matter has a three body decay into $\nu$, $\mu$ and $\tau$ leptons which give rise to a soft spectrum for photon in comparison to the usual two body decay. As a result, bounds from this gamma ray observations~\cite{Cohen2016} do not have one to one correspondence with the present scenario for $\tau\sim 10^{29}$ s and they can be exempted safely.  

It is worth mentioning that, in order to have a long lived PeV scale dark matter candidate along with another long lived particle in order to generate the correct abundance of heavier DM by the mechanism of entropy dilution, we need to fine-tune several parameters of the model. In this minimal setup we have incorporated an approximate $U(1)_X$ global symmetry for this purpose. Another possible way is to go beyond this minimal setup in such a way that the decay of these long lived particles occur only at radiative level, naturally suppressing the decay widths. We leave such a study to future works.

\acknowledgments
UKD acknowledges the support from Department of Science and Technology, Government of India under the fellowship reference number PDF/2016/001087 (SERB National Post-Doctoral Fellowship). DB acknowledges the support from IIT Guwahati start-up grant (reference number: xPHYSUGIITG01152xxDB001) and Associateship Programme of IUCAA, Pune. DB, UKD and GT would like to thank P. S. Bhupal Dev, Michael Gustafsson, Subhendra Mohanty and Ranjan Laha for useful discussions and comments. We would like to thank the organisers of WHEPP XIV at IIT Kanpur where the problem has been conceived.

\bibliographystyle{JHEP}
\bibliography{ref.bib}

\providecommand{\href}[2]{#2}\begingroup\raggedright\begin{thebibliography}{10}

\bibitem{Zwicky:1933gu}
F.~Zwicky, {\it {Die Rotverschiebung von extragalaktischen Nebeln}},  {\em
  Helv. Phys. Acta} {\bf 6} (1933) 110--127. [Gen. Rel. Grav.41,207(2009)].

\bibitem{Rubin:1970zza}
V.~C. Rubin and W.~K. Ford, Jr., {\it {Rotation of the Andromeda Nebula from a
  Spectroscopic Survey of Emission Regions}},  {\em Astrophys. J.} {\bf 159}
  (1970) 379--403.

\bibitem{Clowe:2006eq}
D.~Clowe, M.~Bradac, A.~H. Gonzalez, M.~Markevitch, S.~W. Randall, C.~Jones,
  and D.~Zaritsky, {\it {A direct empirical proof of the existence of dark
  matter}},  {\em Astrophys. J.} {\bf 648} (2006) L109--L113,
  [\href{http://arxiv.org/abs/astro-ph/0608407}{{\tt astro-ph/0608407}}].

\bibitem{Ade:2015xua}
{\bf Planck} Collaboration, P.~A.~R. Ade et~al., {\it {Planck 2015 results.
  XIII. Cosmological parameters}},  {\em Astron. Astrophys.} {\bf 594} (2016)
  A13, [\href{http://arxiv.org/abs/1502.01589}{{\tt arXiv:1502.01589}}].

\bibitem{manalaysay2016}
A.~Manalaysay, {\it Dark-matter results from 332 new live days of lux data},
  July, 2016.

\bibitem{Akerib:2016vxi}
{\bf LUX} Collaboration, D.~S. Akerib et~al., {\it {Results from a search for
  dark matter in the complete LUX exposure}},  {\em Phys. Rev. Lett.} {\bf 118}
  (2017), no.~2 021303, [\href{http://arxiv.org/abs/1608.07648}{{\tt
  arXiv:1608.07648}}].

\bibitem{Tan:2016zwf}
{\bf PandaX-II} Collaboration, A.~Tan et~al., {\it {Dark Matter Results from
  First 98.7 Days of Data from the PandaX-II Experiment}},  {\em Phys. Rev.
  Lett.} {\bf 117} (2016), no.~12 121303,
  [\href{http://arxiv.org/abs/1607.07400}{{\tt arXiv:1607.07400}}].

\bibitem{Aartsen:2013jdh}
{\bf IceCube} Collaboration, M.~G. Aartsen et~al., {\it {Evidence for
  High-Energy Extraterrestrial Neutrinos at the IceCube Detector}},  {\em
  Science} {\bf 342} (2013) 1242856,
  [\href{http://arxiv.org/abs/1311.5238}{{\tt arXiv:1311.5238}}].

\bibitem{Aartsen:2014gkd}
{\bf IceCube} Collaboration, M.~G. Aartsen et~al., {\it {Observation of
  High-Energy Astrophysical Neutrinos in Three Years of IceCube Data}},  {\em
  Phys. Rev. Lett.} {\bf 113} (2014) 101101,
  [\href{http://arxiv.org/abs/1405.5303}{{\tt arXiv:1405.5303}}].

\bibitem{Aartsen:2015zva}
{\bf IceCube} Collaboration, M.~G. Aartsen et~al., {\it {The IceCube Neutrino
  Observatory - Contributions to ICRC 2015 Part II: Atmospheric and
  Astrophysical Diffuse Neutrino Searches of All Flavors}},  in {\em
  {Proceedings, 34th International Cosmic Ray Conference (ICRC 2015): The
  Hague, The Netherlands, July 30-August 6, 2015}}, 2015.
\newblock \href{http://arxiv.org/abs/1510.05223}{{\tt arXiv:1510.05223}}.

\bibitem{Aartsen2016}
{\bf IceCube} Collaboration, M.~G. Aartsen et~al., {\it {Observation and
  Characterization of a Cosmic Muon Neutrino Flux from the Northern Hemisphere
  using six years of IceCube data}},  {\em Astrophys. J.} {\bf 833} (2016),
  no.~1 3, [\href{http://arxiv.org/abs/1607.08006}{{\tt arXiv:1607.08006}}].

\bibitem{Adrian-Martinez:2015ver}
{\bf IceCube, ANTARES} Collaboration, S.~Adrian-Martinez et~al., {\it {The
  First Combined Search for Neutrino Point-sources in the Southern Hemisphere
  With the Antares and Icecube Neutrino Telescopes}},  {\em Astrophys. J.} {\bf
  823} (2016), no.~1 65, [\href{http://arxiv.org/abs/1511.02149}{{\tt
  arXiv:1511.02149}}].

\bibitem{Aartsen:2016tpb}
{\bf IceCube} Collaboration, M.~G. Aartsen et~al., {\it {Lowering IceCube's
  Energy Threshold for Point Source Searches in the Southern Sky}},  {\em
  Astrophys. J.} {\bf 824} (2016), no.~2 L28,
  [\href{http://arxiv.org/abs/1605.00163}{{\tt arXiv:1605.00163}}].

\bibitem{Feldstein:2013kka}
B.~Feldstein, A.~Kusenko, S.~Matsumoto, and T.~T. Yanagida, {\it {Neutrinos at
  IceCube from Heavy Decaying Dark Matter}},  {\em Phys. Rev.} {\bf D88}
  (2013), no.~1 015004, [\href{http://arxiv.org/abs/1303.7320}{{\tt
  arXiv:1303.7320}}].

\bibitem{Esmaili:2013gha}
A.~Esmaili and P.~D. Serpico, {\it {Are IceCube neutrinos unveiling PeV-scale
  decaying dark matter?}},  {\em JCAP} {\bf 1311} (2013) 054,
  [\href{http://arxiv.org/abs/1308.1105}{{\tt arXiv:1308.1105}}].

\bibitem{Ema:2013nda}
Y.~Ema, R.~Jinno, and T.~Moroi, {\it {Cosmic-Ray Neutrinos from the Decay of
  Long-Lived Particle and the Recent IceCube Result}},  {\em Phys. Lett.} {\bf
  B733} (2014) 120--125, [\href{http://arxiv.org/abs/1312.3501}{{\tt
  arXiv:1312.3501}}].

\bibitem{Bhattacharya:2014vwa}
A.~Bhattacharya, M.~H. Reno, and I.~Sarcevic, {\it {Reconciling neutrino flux
  from heavy dark matter decay and recent events at IceCube}},  {\em JHEP} {\bf
  06} (2014) 110, [\href{http://arxiv.org/abs/1403.1862}{{\tt
  arXiv:1403.1862}}].

\bibitem{Rott:2014kfa}
C.~Rott, K.~Kohri, and S.~C. Park, {\it {Superheavy dark matter and IceCube
  neutrino signals: Bounds on decaying dark matter}},  {\em Phys. Rev.} {\bf
  D92} (2015), no.~2 023529, [\href{http://arxiv.org/abs/1408.4575}{{\tt
  arXiv:1408.4575}}].

\bibitem{Esmaili:2014rma}
A.~Esmaili, S.~K. Kang, and P.~D. Serpico, {\it {IceCube events and decaying
  dark matter: hints and constraints}},  {\em JCAP} {\bf 1412} (2014), no.~12
  054, [\href{http://arxiv.org/abs/1410.5979}{{\tt arXiv:1410.5979}}].

\bibitem{Bhattacharya:2014yha}
A.~Bhattacharya, R.~Gandhi, and A.~Gupta, {\it {The Direct Detection of Boosted
  Dark Matter at High Energies and PeV events at IceCube}},  {\em JCAP} {\bf
  1503} (2015), no.~03 027, [\href{http://arxiv.org/abs/1407.3280}{{\tt
  arXiv:1407.3280}}].

\bibitem{Cherry:2014xra}
J.~F. Cherry, A.~Friedland, and I.~M. Shoemaker, {\it {Neutrino Portal Dark
  Matter: From Dwarf Galaxies to IceCube}},
  \href{http://arxiv.org/abs/1411.1071}{{\tt arXiv:1411.1071}}.

\bibitem{Fong:2014bsa}
C.~S. Fong, H.~Minakata, B.~Panes, and R.~Zukanovich~Funchal, {\it {Possible
  Interpretations of IceCube High-Energy Neutrino Events}},  {\em JHEP} {\bf
  02} (2015) 189, [\href{http://arxiv.org/abs/1411.5318}{{\tt
  arXiv:1411.5318}}].

\bibitem{Murase:2015gea}
K.~Murase, R.~Laha, S.~Ando, and M.~Ahlers, {\it {Testing the Dark Matter
  Scenario for PeV Neutrinos Observed in IceCube}},  {\em Phys. Rev. Lett.}
  {\bf 115} (2015), no.~7 071301, [\href{http://arxiv.org/abs/1503.04663}{{\tt
  arXiv:1503.04663}}].

\bibitem{Boucenna:2015tra}
S.~M. Boucenna, M.~Chianese, G.~Mangano, G.~Miele, S.~Morisi, O.~Pisanti, and
  E.~Vitagliano, {\it {Decaying Leptophilic Dark Matter at IceCube}},  {\em
  JCAP} {\bf 1512} (2015), no.~12 055,
  [\href{http://arxiv.org/abs/1507.01000}{{\tt arXiv:1507.01000}}].

\bibitem{Ko:2015nma}
P.~Ko and Y.~Tang, {\it {IceCube Events from Heavy DM decays through the
  Right-handed Neutrino Portal}},  {\em Phys. Lett.} {\bf B751} (2015) 81--88,
  [\href{http://arxiv.org/abs/1508.02500}{{\tt arXiv:1508.02500}}].

\bibitem{ElAisati2015}
C.~El~Aisati, M.~Gustafsson, and T.~Hambye, {\it {New Search for Monochromatic
  Neutrinos from Dark Matter Decay}},  {\em Phys. Rev.} {\bf D92} (2015),
  no.~12 123515, [\href{http://arxiv.org/abs/1506.02657}{{\tt
  arXiv:1506.02657}}].

\bibitem{Dev:2016qbd}
P.~S.~B. Dev, D.~Kazanas, R.~N. Mohapatra, V.~L. Teplitz, and Y.~Zhang, {\it
  {Heavy right-handed neutrino dark matter and PeV neutrinos at IceCube}},
  {\em JCAP} {\bf 1608} (2016), no.~08 034,
  [\href{http://arxiv.org/abs/1606.04517}{{\tt arXiv:1606.04517}}].

\bibitem{Bhattacharya:2016tma}
A.~Bhattacharya, R.~Gandhi, A.~Gupta, and S.~Mukhopadhyay, {\it {Boosted Dark
  Matter and its implications for the features in IceCube HESE data}},  {\em
  JCAP} {\bf 1705} (2017), no.~05 002,
  [\href{http://arxiv.org/abs/1612.02834}{{\tt arXiv:1612.02834}}].

\bibitem{Chianese:2016opp}
M.~Chianese, G.~Miele, S.~Morisi, and E.~Vitagliano, {\it {Low energy IceCube
  data and a possible Dark Matter related excess}},  {\em Phys. Lett.} {\bf
  B757} (2016) 251--256, [\href{http://arxiv.org/abs/1601.02934}{{\tt
  arXiv:1601.02934}}].

\bibitem{DiBari:2016guw}
P.~Di~Bari, P.~O. Ludl, and S.~Palomares-Ruiz, {\it {Unifying leptogenesis,
  dark matter and high-energy neutrinos with right-handed neutrino mixing via
  Higgs portal}},  {\em JCAP} {\bf 1611} (2016), no.~11 044,
  [\href{http://arxiv.org/abs/1606.06238}{{\tt arXiv:1606.06238}}].

\bibitem{Chianese:2016smc}
M.~Chianese and A.~Merle, {\it {A Consistent Theory of Decaying Dark Matter
  Connecting IceCube to the Sesame Street}},
  \href{http://arxiv.org/abs/1607.05283}{{\tt arXiv:1607.05283}}.

\bibitem{Bai:2013nga}
Y.~Bai, R.~Lu, and J.~Salvado, {\it {Geometric Compatibility of IceCube TeV-PeV
  Neutrino Excess and its Galactic Dark Matter Origin}},  {\em JHEP} {\bf 01}
  (2016) 161, [\href{http://arxiv.org/abs/1311.5864}{{\tt arXiv:1311.5864}}].

\bibitem{Higaki:2014dwa}
T.~Higaki, R.~Kitano, and R.~Sato, {\it {Neutrinoful Universe}},  {\em JHEP}
  {\bf 07} (2014) 044, [\href{http://arxiv.org/abs/1405.0013}{{\tt
  arXiv:1405.0013}}].

\bibitem{Ema:2014ufa}
Y.~Ema, R.~Jinno, and T.~Moroi, {\it {Cosmological Implications of High-Energy
  Neutrino Emission from the Decay of Long-Lived Particle}},  {\em JHEP} {\bf
  10} (2014) 150, [\href{http://arxiv.org/abs/1408.1745}{{\tt
  arXiv:1408.1745}}].

\bibitem{Dudas:2014bca}
E.~Dudas, Y.~Mambrini, and K.~A. Olive, {\it {Monochromatic neutrinos generated
  by dark matter and the seesaw mechanism}},  {\em Phys. Rev.} {\bf D91} (2015)
  075001, [\href{http://arxiv.org/abs/1412.3459}{{\tt arXiv:1412.3459}}].

\bibitem{Kopp:2015bfa}
J.~Kopp, J.~Liu, and X.-P. Wang, {\it {Boosted Dark Matter in IceCube and at
  the Galactic Center}},  {\em JHEP} {\bf 04} (2015) 105,
  [\href{http://arxiv.org/abs/1503.02669}{{\tt arXiv:1503.02669}}].

\bibitem{Fiorentin:2016avj}
M.~Re~Fiorentin, V.~Niro, and N.~Fornengo, {\it {A consistent model for
  leptogenesis, dark matter and the IceCube signal}},  {\em JHEP} {\bf 11}
  (2016) 022, [\href{http://arxiv.org/abs/1606.04445}{{\tt arXiv:1606.04445}}].

\bibitem{Bhattacharya:2017jaw}
A.~Bhattacharya, A.~Esmaili, S.~Palomares-Ruiz, and I.~Sarcevic, {\it {Probing
  decaying heavy dark matter with the 4-year IceCube HESE data}},  {\em JCAP}
  {\bf 1707} (2017), no.~07 027, [\href{http://arxiv.org/abs/1706.05746}{{\tt
  arXiv:1706.05746}}].

\bibitem{Pati:1974yy}
J.~C. Pati and A.~Salam, {\it {Lepton Number as the Fourth Color}},  {\em Phys.
  Rev.} {\bf D10} (1974) 275--289. [Erratum: Phys. Rev.D11,703(1975)].

\bibitem{Mohapatra:1974gc}
R.~N. Mohapatra and J.~C. Pati, {\it {A Natural Left-Right Symmetry}},  {\em
  Phys. Rev.} {\bf D11} (1975) 2558.

\bibitem{Senjanovic:1975rk}
G.~Senjanovic and R.~N. Mohapatra, {\it {Exact Left-Right Symmetry and
  Spontaneous Violation of Parity}},  {\em Phys. Rev.} {\bf D12} (1975) 1502.

\bibitem{Mohapatra:1980qe}
R.~N. Mohapatra and R.~E. Marshak, {\it {Local B-L Symmetry of Electroweak
  Interactions, Majorana Neutrinos and Neutron Oscillations}},  {\em Phys. Rev.
  Lett.} {\bf 44} (1980) 1316--1319. [Erratum: Phys. Rev. Lett.44,1643(1980)].

\bibitem{Deshpande:1990ip}
N.~G. Deshpande, J.~F. Gunion, B.~Kayser, and F.~I. Olness, {\it {Left-right
  symmetric electroweak models with triplet Higgs}},  {\em Phys. Rev.} {\bf
  D44} (1991) 837--858.

\bibitem{Mambrini:2015vna}
Y.~Mambrini, N.~Nagata, K.~A. Olive, J.~Quevillon, and J.~Zheng, {\it {Dark
  matter and gauge coupling unification in nonsupersymmetric SO(10) grand
  unified models}},  {\em Phys. Rev.} {\bf D91} (2015), no.~9 095010,
  [\href{http://arxiv.org/abs/1502.06929}{{\tt arXiv:1502.06929}}].

\bibitem{Nagata:2015dma}
N.~Nagata, K.~A. Olive, and J.~Zheng, {\it {Weakly-Interacting Massive
  Particles in Non-supersymmetric SO(10) Grand Unified Models}},  {\em JHEP}
  {\bf 10} (2015) 193, [\href{http://arxiv.org/abs/1509.00809}{{\tt
  arXiv:1509.00809}}].

\bibitem{Borah:2016ees}
D.~Borah, A.~Dasgupta, and S.~Patra, {\it {Common Origin of $3.55$ keV X-ray
  line and Gauge Coupling Unification with Left-Right Dark Matter}},
  \href{http://arxiv.org/abs/1604.01929}{{\tt arXiv:1604.01929}}.

\bibitem{Bandyopadhyay:2017uwc}
T.~Bandyopadhyay and A.~Raychaudhuri, {\it {Left-right model with TeV fermionic
  dark matter and unification}},  \href{http://arxiv.org/abs/1703.08125}{{\tt
  arXiv:1703.08125}}.

\bibitem{Arbelaez:2017ptu}
C.~Arbeláez, M.~Hirsch, and D.~Restrepo, {\it {Fermionic triplet dark matter
  in an $SO(10)$-inspired left right model}},
  \href{http://arxiv.org/abs/1703.08148}{{\tt arXiv:1703.08148}}.

\bibitem{Griest:1989wd}
K.~Griest and M.~Kamionkowski, {\it {Unitarity Limits on the Mass and Radius of
  Dark Matter Particles}},  {\em Phys. Rev. Lett.} {\bf 64} (1990) 615.

\bibitem{Nussinov:2014qva}
S.~Nussinov, {\it {Early inflation induced gravity waves can restrict
  Astro-Particle physics}},  \href{http://arxiv.org/abs/1408.1157}{{\tt
  arXiv:1408.1157}}.

\bibitem{Scherrer:1984fd}
R.~J. Scherrer and M.~S. Turner, {\it {Decaying Particles Do Not Heat Up the
  Universe}},  {\em Phys. Rev.} {\bf D31} (1985) 681.

\bibitem{Valencia:1994cj}
G.~Valencia and S.~Willenbrock, {\it {Quark - lepton unification and rare meson
  decays}},  {\em Phys. Rev.} {\bf D50} (1994) 6843--6848,
  [\href{http://arxiv.org/abs/hep-ph/9409201}{{\tt hep-ph/9409201}}].

\bibitem{Yajnik:2017yhd}
U.~A. Yajnik, {\it {Why PeV scale left-right symmetry is a good thing}},
  \href{http://arxiv.org/abs/1702.03420}{{\tt arXiv:1702.03420}}.

\bibitem{Chang:1983fu}
D.~Chang, R.~N. Mohapatra, and M.~K. Parida, {\it {Decoupling Parity and
  SU(2)-R Breaking Scales: A New Approach to Left-Right Symmetric Models}},
  {\em Phys. Rev. Lett.} {\bf 52} (1984) 1072.

\bibitem{Chang:1984uy}
D.~Chang, R.~N. Mohapatra, and M.~K. Parida, {\it {A New Approach to Left-Right
  Symmetry Breaking in Unified Gauge Theories}},  {\em Phys. Rev.} {\bf D30}
  (1984) 1052.

\bibitem{Chang:1984qr}
D.~Chang, R.~N. Mohapatra, J.~Gipson, R.~E. Marshak, and M.~K. Parida, {\it
  {Experimental Tests of New SO(10) Grand Unification}},  {\em Phys. Rev.} {\bf
  D31} (1985) 1718.

\bibitem{ATLAS:2012ak}
{\bf ATLAS} Collaboration, G.~Aad et~al., {\it {Search for heavy neutrinos and
  right-handed $W$ bosons in events with two leptons and jets in $pp$
  collisions at $\sqrt{s}=7$ TeV with the ATLAS detector}},  {\em Eur. Phys.
  J.} {\bf C72} (2012) 2056, [\href{http://arxiv.org/abs/1203.5420}{{\tt
  arXiv:1203.5420}}].

\bibitem{CMS:2012zv}
{\bf CMS} Collaboration, S.~Chatrchyan et~al., {\it {Search for heavy neutrinos
  and W[R] bosons with right-handed couplings in a left-right symmetric model
  in pp collisions at sqrt(s) = 7 TeV}},  {\em Phys. Rev. Lett.} {\bf 109}
  (2012) 261802, [\href{http://arxiv.org/abs/1210.2402}{{\tt
  arXiv:1210.2402}}].

\bibitem{Heeck:2015qra}
J.~Heeck and S.~Patra, {\it {Minimal Left-Right Symmetric Dark Matter}},  {\em
  Phys. Rev. Lett.} {\bf 115} (2015), no.~12 121804,
  [\href{http://arxiv.org/abs/1507.01584}{{\tt arXiv:1507.01584}}].

\bibitem{Patra:2014goa}
S.~Patra and P.~Pritimita, {\it {Post-sphaleron baryogenesis and $n$ -
  $\overline{n}$ oscillation in non-SUSY SO(10) GUT with gauge coupling
  unification and proton decay}},  {\em Eur. Phys. J.} {\bf C74} (2014), no.~10
  3078, [\href{http://arxiv.org/abs/1405.6836}{{\tt arXiv:1405.6836}}].

\bibitem{Deppisch:2014qpa}
F.~F. Deppisch, T.~E. Gonzalo, S.~Patra, N.~Sahu, and U.~Sarkar, {\it {Signal
  of Right-Handed Charged Gauge Bosons at the LHC?}},  {\em Phys. Rev.} {\bf
  D90} (2014), no.~5 053014, [\href{http://arxiv.org/abs/1407.5384}{{\tt
  arXiv:1407.5384}}].

\bibitem{Deppisch:2014zta}
F.~F. Deppisch, T.~E. Gonzalo, S.~Patra, N.~Sahu, and U.~Sarkar, {\it {Double
  beta decay, lepton flavor violation, and collider signatures of left-right
  symmetric models with spontaneous $D$-parity breaking}},  {\em Phys. Rev.}
  {\bf D91} (2015), no.~1 015018, [\href{http://arxiv.org/abs/1410.6427}{{\tt
  arXiv:1410.6427}}].

\bibitem{Hati:2017aez}
C.~Hati, S.~Patra, M.~Reig, J.~W.~F. Valle, and C.~A. Vaquera-Araujo, {\it
  {Towards gauge coupling unification in left-right symmetric $\mathrm{SU(3)_c
  \times SU(3)_L \times SU(3)_R \times U(1)_{X}}$ theories}},
  \href{http://arxiv.org/abs/1703.09647}{{\tt arXiv:1703.09647}}.

\bibitem{Deppisch:2016scs}
F.~F. Deppisch, C.~Hati, S.~Patra, P.~Pritimita, and U.~Sarkar, {\it
  {Implications of the diphoton excess on left–right models and gauge
  unification}},  {\em Phys. Lett.} {\bf B757} (2016) 223--230,
  [\href{http://arxiv.org/abs/1601.00952}{{\tt arXiv:1601.00952}}].

\bibitem{Patra:2015bga}
S.~Patra, F.~S. Queiroz, and W.~Rodejohann, {\it {Stringent Dilepton Bounds on
  Left-Right Models using LHC data}},  {\em Phys. Lett.} {\bf B752} (2016)
  186--190, [\href{http://arxiv.org/abs/1506.03456}{{\tt arXiv:1506.03456}}].

\bibitem{Awasthi:2013ff}
R.~L. Awasthi, M.~K. Parida, and S.~Patra, {\it {Neutrino masses, dominant
  neutrinoless double beta decay, and observable lepton flavor violation in
  left-right models and SO(10) grand unification with low mass $ W_R, Z_R$
  bosons}},  {\em JHEP} {\bf 08} (2013) 122,
  [\href{http://arxiv.org/abs/1302.0672}{{\tt arXiv:1302.0672}}].

\bibitem{Borah:2013lva}
D.~Borah, S.~Patra, and P.~Pritimita, {\it {Sub-dominant type-II seesaw as an
  origin of non-zero $\theta_{13}$ in SO(10) model with TeV scale Z' gauge
  boson}},  {\em Nucl. Phys.} {\bf B881} (2014) 444--466,
  [\href{http://arxiv.org/abs/1312.5885}{{\tt arXiv:1312.5885}}].

\bibitem{Dev:2015pga}
P.~S. Bhupal~Dev and R.~N. Mohapatra, {\it {Unified explanation of the $eejj$,
  diboson and dijet resonances at the LHC}},  {\em Phys. Rev. Lett.} {\bf 115}
  (2015), no.~18 181803, [\href{http://arxiv.org/abs/1508.02277}{{\tt
  arXiv:1508.02277}}].

\bibitem{Jungman:1995df}
G.~Jungman, M.~Kamionkowski, and K.~Griest, {\it {Supersymmetric dark matter}},
   {\em Phys. Rept.} {\bf 267} (1996) 195--373,
  [\href{http://arxiv.org/abs/hep-ph/9506380}{{\tt hep-ph/9506380}}].

\bibitem{Garcia-Cely:2015quu}
C.~Garcia-Cely and J.~Heeck, {\it {Phenomenology of left-right symmetric dark
  matter}},  \href{http://arxiv.org/abs/1512.03332}{{\tt arXiv:1512.03332}}.
  [JCAP1603,021(2016)].

\bibitem{Ibe:2008ye}
M.~Ibe, H.~Murayama, and T.~T. Yanagida, {\it {Breit-Wigner Enhancement of Dark
  Matter Annihilation}},  {\em Phys. Rev.} {\bf D79} (2009) 095009,
  [\href{http://arxiv.org/abs/0812.0072}{{\tt arXiv:0812.0072}}].

\bibitem{Kolb:1990vq}
E.~W. Kolb and M.~S. Turner, {\it {The Early Universe}},  {\em Front. Phys.}
  {\bf 69} (1990) 1--547.

\bibitem{Scherrer:1985zt}
R.~J. Scherrer and M.~S. Turner, {\it {On the Relic, Cosmic Abundance of Stable
  Weakly Interacting Massive Particles}},  {\em Phys. Rev.} {\bf D33} (1986)
  1585. [Erratum: Phys. Rev.D34,3263(1986)].

\bibitem{Gondolo:1990dk}
P.~Gondolo and G.~Gelmini, {\it {Cosmic abundances of stable particles:
  Improved analysis}},  {\em Nucl. Phys.} {\bf B360} (1991) 145--179.

\bibitem{Griest:1990kh}
K.~Griest and D.~Seckel, {\it {Three exceptions in the calculation of relic
  abundances}},  {\em Phys. Rev.} {\bf D43} (1991) 3191--3203.

\bibitem{Dev:2016xcp}
P.~S. Bhupal~Dev, R.~N. Mohapatra, and Y.~Zhang, {\it {Naturally stable
  right-handed neutrino dark matter}},  {\em JHEP} {\bf 11} (2016) 077,
  [\href{http://arxiv.org/abs/1608.06266}{{\tt arXiv:1608.06266}}].

\bibitem{Dev:2016qeb}
P.~S.~B. Dev, R.~N. Mohapatra, and Y.~Zhang, {\it {Heavy right-handed neutrino
  dark matter in left-right models}},
  \href{http://arxiv.org/abs/1610.05738}{{\tt arXiv:1610.05738}}.

\bibitem{Agashe:2014kda}
{\bf Particle Data Group} Collaboration, K.~A. Olive et~al., {\it {Review of
  Particle Physics}},  {\em Chin. Phys.} {\bf C38} (2014) 090001.

\bibitem{Vincent:2016nut}
A.~C. Vincent, S.~Palomares-Ruiz, and O.~Mena, {\it {Analysis of the 4-year
  IceCube high-energy starting events}},  {\em Phys. Rev.} {\bf D94} (2016),
  no.~2 023009, [\href{http://arxiv.org/abs/1605.01556}{{\tt
  arXiv:1605.01556}}].

\bibitem{Kistler2016}
M.~D. Kistler and R.~Laha, {\it {Multi-PeV Signals from a New Astrophysical
  Neutrino Flux Beyond the Glashow Resonance}},
  \href{http://arxiv.org/abs/1605.08781}{{\tt arXiv:1605.08781}}.

\bibitem{Anchordoqui2014}
L.~A. Anchordoqui, V.~Barger, H.~Goldberg, J.~G. Learned, D.~Marfatia,
  S.~Pakvasa, T.~C. Paul, and T.~J. Weiler, {\it {End of the cosmic neutrino
  energy spectrum}},  {\em Phys. Lett.} {\bf B739} (2014) 99--101,
  [\href{http://arxiv.org/abs/1404.0622}{{\tt arXiv:1404.0622}}].

\bibitem{Learned2014}
J.~G. Learned and T.~J. Weiler, {\it {A Relational Argument for a $\sim$PeV
  Neutrino Energy Cutoff}},  \href{http://arxiv.org/abs/1407.0739}{{\tt
  arXiv:1407.0739}}.

\bibitem{Tomar2015}
G.~Tomar, S.~Mohanty, and S.~Pakvasa, {\it {Lorentz Invariance Violation and
  IceCube Neutrino Events}},  {\em JHEP} {\bf 11} (2015) 022,
  [\href{http://arxiv.org/abs/1507.03193}{{\tt arXiv:1507.03193}}].

\bibitem{Esmaili:2012us}
A.~Esmaili, A.~Ibarra, and O.~L.~G. Peres, {\it {Probing the stability of
  superheavy dark matter particles with high-energy neutrinos}},  {\em JCAP}
  {\bf 1211} (2012) 034, [\href{http://arxiv.org/abs/1205.5281}{{\tt
  arXiv:1205.5281}}].

\bibitem{Cirelli2011}
M.~Cirelli, G.~Corcella, A.~Hektor, G.~Hutsi, M.~Kadastik, P.~Panci, M.~Raidal,
  F.~Sala, and A.~Strumia, {\it {PPPC 4 DM ID: A Poor Particle Physicist
  Cookbook for Dark Matter Indirect Detection}},  {\em JCAP} {\bf 1103} (2011)
  051, [\href{http://arxiv.org/abs/1012.4515}{{\tt arXiv:1012.4515}}].
  [Erratum: JCAP1210,E01(2012)].

\bibitem{Cline2013}
J.~M. Cline and P.~Scott, {\it {Dark Matter CMB Constraints and Likelihoods for
  Poor Particle Physicists}},  {\em JCAP} {\bf 1303} (2013) 044,
  [\href{http://arxiv.org/abs/1301.5908}{{\tt arXiv:1301.5908}}]. [Erratum:
  JCAP1305,E01(2013)].

\bibitem{Navarro:1996gj}
J.~F. Navarro, C.~S. Frenk, and S.~D.~M. White, {\it {A Universal density
  profile from hierarchical clustering}},  {\em Astrophys. J.} {\bf 490} (1997)
  493--508, [\href{http://arxiv.org/abs/astro-ph/9611107}{{\tt
  astro-ph/9611107}}].

\bibitem{Chakraborty2015}
S.~Chakraborty and I.~Izaguirre, {\it {Diffuse neutrinos from extragalactic
  supernova remnants: Dominating the 100 TeV IceCube flux}},  {\em Phys. Lett.}
  {\bf B745} (2015) 35--39, [\href{http://arxiv.org/abs/1501.02615}{{\tt
  arXiv:1501.02615}}].

\bibitem{Stecker1991}
F.~W. Stecker, C.~Done, M.~H. Salamon, and P.~Sommers, {\it {High-energy
  neutrinos from active galactic nuclei}},  {\em Phys. Rev. Lett.} {\bf 66}
  (1991) 2697--2700. [Erratum: Phys. Rev. Lett.69,2738(1992)].

\bibitem{Kalashev2015}
O.~Kalashev, D.~Semikoz, and I.~Tkachev, {\it {Neutrinos in IceCube from active
  galactic nuclei}},  {\em J. Exp. Theor. Phys.} {\bf 120} (2015), no.~3
  541--548, [\href{http://arxiv.org/abs/1410.8124}{{\tt arXiv:1410.8124}}].

\bibitem{Waxman1997}
E.~Waxman and J.~N. Bahcall, {\it {High-energy neutrinos from cosmological
  gamma-ray burst fireballs}},  {\em Phys. Rev. Lett.} {\bf 78} (1997)
  2292--2295, [\href{http://arxiv.org/abs/astro-ph/9701231}{{\tt
  astro-ph/9701231}}].

\bibitem{Barger2014}
V.~Barger, L.~Fu, J.~G. Learned, D.~Marfatia, S.~Pakvasa, and T.~J. Weiler,
  {\it {Glashow resonance as a window into cosmic neutrino sources}},  {\em
  Phys. Rev.} {\bf D90} (2014) 121301,
  [\href{http://arxiv.org/abs/1407.3255}{{\tt arXiv:1407.3255}}].

\bibitem{Chen:2014gxa}
C.-Y. Chen, P.~S. Bhupal~Dev, and A.~Soni, {\it {Two-component flux explanation
  for the high energy neutrino events at IceCube}},  {\em Phys. Rev.} {\bf D92}
  (2015), no.~7 073001, [\href{http://arxiv.org/abs/1411.5658}{{\tt
  arXiv:1411.5658}}].

\bibitem{Palomares-Ruiz:2015mka}
S.~Palomares-Ruiz, A.~C. Vincent, and O.~Mena, {\it {Spectral analysis of the
  high-energy IceCube neutrinos}},  {\em Phys. Rev.} {\bf D91} (2015), no.~10
  103008, [\href{http://arxiv.org/abs/1502.02649}{{\tt arXiv:1502.02649}}].

\bibitem{Cohen2016}
T.~Cohen, K.~Murase, N.~L. Rodd, B.~R. Safdi, and Y.~Soreq, {\it {Gamma-ray
  Constraints on Decaying Dark Matter and Implications for IceCube}},
  \href{http://arxiv.org/abs/1612.05638}{{\tt arXiv:1612.05638}}.

\end{thebibliography}\endgroup

\end{document}